\documentstyle[12pt,epsf]{article} 
\begin{document}
\baselineskip 14pt plus 2pt minus 2pt
\newcommand{\beq}{\begin{equation}}
\newcommand{\eeq}{\end{equation}}
\newcommand{\beqa}{\begin{eqnarray}}
\newcommand{\eeqa}{\end{eqnarray}}
\newcommand{\dfrac}{\displaystyle \frac}
\renewcommand{\thefootnote}{\#\arabic{footnote}}
\newcommand{\ve}{\varepsilon}
\newcommand{\krig}[1]{\stackrel{\circ}{#1}}
\newcommand{\barr}[1]{\not\mathrel #1}
\newcommand{\vs}{\vspace{-0.25cm}}

\begin{titlepage}
 
\noindent FZJ-IKP(TH)-1998-03
\hfill {\tt hep-ph/9803266}

\vspace{1.0cm}

\begin{center}

{\large  \bf {
Pion--nucleon scattering in chiral perturbation theory I:\\[0.2em]
Isospin--symmetric case}}\footnote{Work supported
    in part by funds provided by the Graduiertenkolleg "Die Erforschung 
    subnuklearer Strukturen der Materie" at Bonn University.}

\vspace{1.2cm}
                              
{\large Nadia Fettes$^{\ddag}$\footnote{email: N.Fettes@fz-juelich.de},
Ulf-G. Mei\ss ner$^{\ddag}$\footnote{email: Ulf-G.Meissner@fz-juelich.de},
Sven Steininger$^{\ddag,\dag}$\footnote{email: sven@pythia.itkp.uni-bonn.de}
}

\vspace{1.0cm}

$^{\ddag}$FZ J\"ulich, IKP (Theorie), D--52425 J\"ulich, Germany\\

\vspace{0.4cm}
$^{\dag}$Universit\"at Bonn, ITKP, 
Nussallee 14-16, D--53115 Bonn, Germany\\

\end{center}

\vspace{0.8cm}

\begin{abstract}
\noindent We construct the complete  effective chiral pion--nucleon
Lagrangian to third order in small momenta based on relativistic
chiral perturbation theory. We then perform the so--called heavy
baryon limit and construct all terms up--to--and--including order 
$1/m^2$ with fixed and
free coefficients. As an application, we discuss in detail pion--nucleon
scattering. In particular, we show that for this process and to third
order, the $1/m$ expansion of the
Born graphs calculated relativistically can be recovered exactly  in
the heavy baryon approach without any additional momentum--dependent
wave function renormalization. We fit various empirical phase shifts
for pion  laboratory momenta between 50 and 100~MeV.
This leads to a satisfactory description of the phase shifts up
to momenta of about 200~MeV. We also predict the threshold parameters,
which turn out to be in good agreement with the dispersive analysis. 
In particular, we can sharpen the prediction for the isovector S--wave
scattering length, $0.083\, M_\pi^{-1} \leq a_{0+}^- \leq 0.093\, 
M_\pi^{-1}$. We also consider the subthreshold parameters and
give a short comparison to other calculations of $\pi N$ scattering 
in chiral perturbation theory or modifications thereof.

\end{abstract}


\vspace{2cm}

\centerline{Accepted for publication in Nuclear Physics A}

\vfill

\end{titlepage}

\section{Introduction and summary}
\label{sec:intro}
\def\theequation{\arabic{section}.\arabic{equation}}
\setcounter{equation}{0}

One of the outstanding problems in nuclear and particle physics is 
to understand quantitatively the violation of isospin symmetry. In a world
of strong interactions only, the light quark mass difference $m_d-m_u$
is the only mechanism for isospin violation. In the real world, the
situation is complicated by the electromagnetic interactions, which clearly
do not conserve isospin (the photon couples to the electric charge).
With the advent of high--precision data on pionic hydrogen and deuterium at 
PSI~\cite{sigg} and neutral pion photoproduction data from
MAMI~\cite{fuchs} and SAL~\cite{berg}, novel interest has been spurred
in separating electromagnetic effects and trying to filter out isospin
violating contributions. In addition, in the framework of some models
it has been claimed that the presently available pion--nucleon data
basis exhibits isospin violation of the order of a few 
percent~\cite{gibbs}\cite{matsi}. However, to really pin down isospin
breaking due to the light quark mass difference, one needs a machinery
that allows to {\it simultaneously} and {\it consistently} 
treat the electromagnetic and the
strong contributions. The only framework known at present allowing to
do just that is (heavy baryon) chiral perturbation theory (HBCHPT). 
It is based on a consistent power counting
scheme which allows to order the various terms according to the number
of derivatives and/or meson mass insertions. This defines the
so--called chiral dimension corresponding to the number
of derivatives and/or pion mass insertions, with the pertinent
small parameters collectively denoted by $q$. For the pion--nucleon
system, the leading term is of dimension one and loops start to
contribute at two orders higher (in the heavy fermion formulation).
HBCHPT has become a precision tool to
investigate a rich variety of  processes in the pion--nucleon system, in
particular reactions involving real and virtual photons (for a review,
see~\cite{bkmrev} and an update is given in~\cite{ulflec}).

Over the years, Bernard et al. have  investigated aspects of 
pion--nucleon scattering,
in particular the chiral corrections to the S--wave scattering 
lengths~\cite{bkmpin,bkmpin2}, the size of the remainder at the Cheng-Dashen
point~\cite{bkmcd} and the strength of the dimension two low--energy
constants (LECs) extracted from $\pi N$ threshold and subthreshold parameters,
that do not depend on dimension three LECs~\cite{bkmlec}. Most of these
calculations were done in the one--loop approximation, some to order $q^3$
and some to order $q^4$.\footnote{Note that due to the appearance of odd and
even powers in the effective pion--nucleon Lagrangian, the $N$--loop order 
consists of terms of order $q^{2N+1}$ and $q^{2N+2}$.} 
The complete one--loop amplitude to order $q^3$ was first 
given by Moj\v zi\v s~\cite{moj} based on the effective Lagrangian constructed 
in~\cite{eckmoj}. In ref.\cite{moj}, the Karlsruhe--Helsinki~\cite{kopi}
threshold parameters together with the pion--nucleon coupling constant
and the $\pi N$ $\sigma$--term were used to pin down the various LECs of
dimension two and three. The resulting phase shifts were later evaluated
for pion kinetic laboratory energies up to 140~MeV in~\cite{dapa}. Simultaneously,
the effective Lagrangian including also virtual photons was developed in~\cite{ms}
to third order in small momenta (counting also the electric charge $e$
as an  additional
small parameter) and isospin--violating effects for the scattering of neutral pions off
nucleons and in the $\sigma$--term were evaluated. This extended effective
Lagrangian was based on the pion--nucleon Lagrangian as formulated in the
review~\cite{bkmrev}. To allow for an easier comparison with the various
calculations performed in that basis, we construct  here the corresponding
third order effective Lagrangian. This extends the work of Ecker and 
Moj\v zi\v s~\cite{eckmoj} since we explicitely work out the various $1/m$
corrections to the dimension three terms and do not absorb them in the
corresponding LECs (as done in \cite{eckmoj}). We also fill in some details
on the construction of the effective Lagrangian which have not yet appeared
in the literature. In the second part of this
investigation, we use pion--nucleon phase shifts in the low--energy
region as given by different analyses to pin down our LECs. We argue that
considering the present status in the pion--nucleon scattering data basis,
this is a more stable procedure than the one used before. In particular, it
allows to extract the theoretical errors related to the variations in the
empirical input. This is of utmost importance if one intends to continue
these studies to next order and work out the size and origin of the isospin
violation.
 
The pertinent results of this investigation can be summarized as follows:
\begin{enumerate}
\item[(i)] We have constructed the minimal pion--nucleon
Lagrangian at third order in the chiral expansion, including all corrections
arising from the expansion in the inverse nucleon mass (with fixed coefficients
as well as dimension two LECs). We have also enumerated the purely divergent
terms which arise from the evaluation of the one--loop generating functional
to one loop and order $q^3$. At third order, the effective Lagrangian
has 23 terms with LECs. Only 14 of these have a non--vanishing $\beta$--function.
This extends the result obtained in~\cite{eckmoj}.
\smallskip
\item[(ii)] We have constructed the complete  amplitude for 
elastic pion--nucleon scattering to order $q^3$. 
Our loop contribution agrees exactly with
the one given previously~\cite{moj,bkmlec}. For the tree graphs, we have
shown the exact equivalence between the $1/m$ expansion of the result obtained
in the relativistic theory and the one calculated directly in the heavy baryon
approach, provided one does not expand the normalization factor of the nucleon
spinors.
\smallskip
\item[(iii)] We have fitted the two S-- and four P--wave partial wave amplitudes
for three different sets of available pion--nucleon phase shifts at intermediate
energies (typically in the range of 50 to 100~MeV pion momentum in the
laboratory frame). This allows to predict the phases at {\it lower} and at 
{\it higher} energies, in particular the (sub)threshold parameters (scattering
lengths and effective ranges). By this fitting procedure we can determine
the four dimension two LECs $c_{1,2,3,4}$ and five (combinations) of dimension
three LECs. The numerical values for the LECs $c_{2,3,4}$ are in good
agreement with previous determinations from threshold and subthreshold parameters
alone~\cite{bkmlec,moj}. Consequently, the numerical values of these LECs
can be understood by resonance exchange saturation, see~\cite{bkmlec}.
The LEC $c_1$, which is directly related to the $\sigma$--term, comes
out larger than before. We show, however, that the fits are not very sensitive
to its actual value and one  thus can not pin down the $\sigma$--term at this
order in the chiral expansion. The dimension three LECs have ``natural'' size.
\smallskip
\item[(iv)] We have evaluated the threshold parameters using the one--loop
amplitude for the extrapolation. The results are fairly stable for the
various fits and, if applicable, comparable to the results based on dispersion
theory. As already known, the isoscalar S--wave scattering length can not
be determined precisely to this order, but we give an improved prediction
for the isovector one, $0.083\, M_\pi^{-1} \leq a_{0+}^- \leq 0.093\, 
M_\pi^{-1}$. This value is in good agreement with recent determinations from the
strong interaction level shift in pionic atoms measured at
PSI~\cite{sigg} and with values extracted from phase shift analyses.
\end{enumerate}

\section{Construction of the effective Lagrangian}
\setcounter{equation}{0}

In this section, we give a detailed exposition how to arrive at the
dimension three effective chiral pion--nucleon Lagrangian. While
parts of the final result have already been given by Ecker and Moj\v zi\v s
\cite{eckmoj}, we fill in the necessary details and also work in the
basis of the review~\cite{bkmrev}, which is most convenient for
comparison with earlier work on photoproduction, Compton scattering
and so on. In particular, in \cite{eckmoj} the $1/m$ corrections
appearing at third order were always subsumed in the LECs (whenever
possible). While this is a legitimate procedure, we prefer these
corrections to be separated, in particular, if one wishes to estimate
the corresponding LECs via resonance exchange or some other model.
The construction of ${\cal L}_{\pi N}^{(3)}$ 
is done in five steps. These are: (1) enumeration of the building
blocks for relativistic spin--1/2 fields chirally coupled to pions and
external sources, (2) saturation of the free Lorentz indices by elements of
the underlying Clifford algebra and other operators of chiral dimension
zero (and one), (3) construction of the overcomplete relativistic Lagrangian,
(4) reduction of terms by use of various relations and (5) performing
the non--relativistic limit and working out the $1/m$ corrections by
use of the path integral~\cite{bkkm}. The details will be spelled out
in the following paragraphs.\footnote{After submission of this paper,
another manuscript which also deals with this topic appeared~\cite{baur}.}
The reader more interested in the application to
pion--nucleon scattering might skip this section.

\subsection{Effective theory with relativistic nucleons}

The starting point of our approach is to construct the most general
chiral invariant Lagrangian built from pions, nucleons and external
scalar, pseudoscalar, vector and axial--vector sources. The Goldstone
bosons are collected in a 2$\times$2 matrix-valued field $U(x)= u^2(x)$.
The nucleons are described by structureless relativistic spin-${\small
  \frac{1}{2}}$ particles, the spinors being denoted by $\Psi
(x)$. The effective theory admits a low energy expansion, i.e. the
corresponding effective Lagrangian can be written as (for more details
and references, see e.g. \cite{bkmrev})
\beq
{\cal L}_{\rm eff} = 
 {\cal L}_{\pi\pi}^{(2)} +  {\cal L}_{\pi\pi}^{(4)} +
 {\cal L}_{\pi N}^{(1)} +  {\cal L}_{\pi N}^{(2)} +
 {\cal L}_{\pi N}^{(3)} + \ldots~,
\eeq
where the ellipsis denotes terms of higher order. For the explicit
form of the meson Lagrangian and the dimension one and two
pion--nucleon terms, we refer to ref.\cite{bkmrev}. We remark that
to be precise, the various parameters like $g_A, m, \ldots$ 
appearing in the effective
Lagrangian have to be taken in the chiral SU(2) limit ($m_u=m_d=0\, ,
\,\, m_s$ fixed) and should be denoted as $\krig{g}_A , \krig{m},
\ldots$. Throughout this section, we will not specify this but it
should be kept in mind.  In what follows, we are mostly
interested in the third order terms collected in ${\cal L}_{\pi N}^{(3)}$.
To construct these, we use the standard methods of non--linearly
realized chiral symmetry. From the external sources, we construct the
following building blocks starting with the covariant derivative 
$D_\mu$ (here and in what follows, $A$ denotes an arbitrary 2$\times$2
matrix):
\beqa
D_\mu A & = & \partial_\mu\,A + \Gamma_\mu\,A~, \\ 
\Gamma_\mu & = & \frac{1}{2}\,[u^\dagger,\partial_\mu u] 
                 - \frac{i}{2}\,u^\dagger r_\mu u
                 - \frac{i}{2}\,u l_\mu u^\dagger ~,\\
u_\mu & = & i\,\left(\partial_\mu u\,u^\dagger + u^\dagger\partial_\mu u 
            -i\,u^\dagger r_\mu u + i\,u l_\mu u^\dagger\right)~, \\
r_\mu & = & v_\mu + a_\mu~, \quad l_\mu  =  v_\mu - a_\mu~, \\
F^\pm_{\mu\nu} & = & u^\dagger F^R_{\mu\nu}u \pm u F^L_{\mu\nu}u^\dagger~,\\
F^L_{\mu\nu} & = & \partial_\mu l_\nu - \partial_\nu l_\mu - i\,[l_\mu,l_\nu]~, \\
F^R_{\mu\nu} & = & \partial_\mu r_\nu - \partial_\nu r_\mu - i\,[r_\mu,r_\nu]~, \\
\chi_\pm & = & u^\dagger \chi u^\dagger \pm u \chi^\dagger u~,
\eeqa
and we employ the following definition for traceless operators
('$\langle \ldots \rangle$' denotes the trace in flavor space)
\beq
\hat{A}  =  A - \frac{1}{2}\,\langle A \rangle~. 
\eeq
Here, $\chi (x)= s(x) + ip(x)$ includes the explicit chiral symmetry breaking
through the small current quark masses, $s(x)= B \, {\rm diag}(m_u,m_d) +
\ldots$ and
\beq
B = \frac{|\langle 0 | \bar q q | 0 \rangle|}{F_\pi^2}~,
\eeq
with $F_\pi$ the (weak) pion decay constant and the scalar quark
condensate is the order parameter of the chiral symmetry breaking.
We assign the following chiral dimensions to the pertinent
fields and operators:
\beqa
&& U(x) , \Psi (x) ,   \partial_\mu \Psi (x) = {\cal O}(1)~, \nonumber \\
&& \partial _\mu U(x) , u_\mu (x) , 
l_\mu (x) , r_\mu (x) = {\cal O}(q)~,\\
&& s(x), p(x), \chi_\pm , F_{\mu\nu}^\pm (x) ,   F_{\mu\nu}^{L,R} (x)  
= {\cal O}(q^2)~,\nonumber
\eeqa
which amounts to the so--called large condensate scenario of the
spontaneous chiral symmetry violation, $B \gg F_\pi$.\footnote{For the
small $B$ scenario, one would have to count $s(x)$ and $p(x)$ as order $q$.}
Here, $q$ denotes a genuine small momentum or meson mass with respect
to the typical hadronic scale of about  1~GeV.
Furthermore, in the following we will make use of the operator relations
\beqa\label{Dcomm}
[D_\mu,D_\nu] & = & \frac{1}{4}\,[u_\mu,u_\nu] - \frac{i}{2}\, F^+_{\mu\nu}~, \\
F^-_{\mu\nu} & = & [D_\mu,u_\nu] - [D_\nu,u_\mu]~. 
\eeqa
Note that if we are only dealing with photons ($A_\mu$) as external fields,
$F^\pm_{\mu\nu}$ simplifies to $F^\pm_{\mu\nu} = (\partial_\mu A_\nu -
\partial_\nu A_\mu)(uQu^\dagger \pm u^\dagger Qu)$ with $Q = e \, {\rm
  diag}(1,0)$ the nucleon charge matrix. For a consistent counting
scheme including also virtual photons~\cite{mms}, one has to consider
the electric charge $e$ as another small quantity of order $q$.

There is one subtlety in the fully relativistic theory, namely the
appearance of the large nucleon mass scale (to be precise, the mass
of the nucleon in the chiral limit)~\cite{gss}. Therefore
any covariant derivative $D_\mu$ counts as order $q^0$. Formally, one
could thus construct terms of the type $\bar \Psi \ldots
D^{38}\Psi$. To avoid this, consider the covariant derivative only as 
a building block when it acts on operators sandwiched between the
spinors (the only exception to this rule is the lowest order
Lagrangian). The pertinent terms involving $D_\mu$ acting on the
nucleon fields, which are all of chiral dimension zero (or one), are given
together with the elements of the Clifford algebra in the next paragraph since
they are used to saturate the free Lorentz indices of the building
blocks. In this way, one avoids from the beginning all terms with an
arbitrary large number of $D_\mu$'s, which are formally allowed. With
that in mind, the possible building blocks at orders $q$, $q^2$ and
$q^3$ are enumerated in table~1 (also given are the respective 
charge conjugation, $C$, and parity, $P$, assignments).
\renewcommand{\arraystretch}{1.2}
\begin{table}[hbt]
\begin{eqnarray*}
\begin{array}{|c|c|c|c|c|c|c|}
\hline
\mbox{Operator} & \quad C \quad & \quad P \quad & \!\!\!\!\!\!\!& 
\mbox{Operator} & \quad C \quad & \quad P \quad \\[0.05em] \hline 
\hline
& & & & & & \\[-1.4em]
u_\mu    & +  & - & &
[D_\mu, \langle\chi_+\rangle ] & + & +\\[0.05em] \hline  
& & & & & & \\[-1.4em]
& {\cal O}(q^2) & & &
\langle\hat{\chi}_- u_\mu\rangle & + & + \\[0.05em] \hline
& & & & & & \\[-1.4em]
\hat{\chi}_+ & + & + & &
\langle\chi_-\rangle u_\mu & + & + \\[0.05em] \hline
& & & & & & \\[-1.4em]
\langle\chi_+\rangle & + & + & &
[\chi_-,u_\mu] & - & + \\[0.05em] \hline
& & & & & & \\[-1.4em]
\hat{\chi}_- & + & - & &
[D_\mu, \hat{\chi}_-] & + & - \\[0.05em] \hline
& & & & & & \\[-1.4em]
\langle\chi_-\rangle & + & - & &
[D_\mu, \langle\chi_-\rangle ]  & + & - \\[0.05em] \hline
& & & & & & \\[-1.4em]
\langle u_\mu u_\nu \rangle & + & + & &
\langle u_\mu u_\nu\rangle u_\lambda & + & - \\[0.05em] \hline
& & & & & & \\[-1.4em]
[u_\mu,u_\nu] & - & + & &
\langle u_\mu [u_\nu,u_\lambda]\rangle & - & - \\[0.05em] \hline 
& & & & & & \\[-1.4em]
[D_\mu,u_\nu] & + & - & &
\langle [D_\mu,u_\nu] u_\lambda\rangle & + & + \\[0.05em] \hline
& & & & & & \\[-1.4em]
\hat{F}^+_{\mu\nu} & - & + & &
[[D_\mu,u_\nu],u_\lambda] & - & + \\[0.05em] \hline
& & & & & & \\[-1.4em]
\langle F^+_{\mu\nu}\rangle & - & + & &
[D_\mu,[D_\nu,u_\lambda]] & + & - \\[0.05em] \hline
& & & & & & \\[-1.4em]
& {\cal O}(q^3) & & &
\langle\hat{F}^+_{\mu\nu} u_\lambda\rangle & - & - \\[0.05em] \hline
& & & & & & \\[-1.4em]
\langle\hat{\chi}_+ u_\mu\rangle & + & - & &
\langle F^+_{\mu\nu}\rangle u_\lambda & - & - \\[0.05em] \hline
& & & & & & \\[-1.4em]
\langle\chi_+\rangle u_\mu & + & - & &
[\hat{F}^+_{\mu\nu},u_\lambda] & + & - \\[0.05em] \hline
& & & & & & \\[-1.4em]
[\chi_+,u_\mu] & - & - & &
[D_\lambda,\hat{F}^+_{\mu\nu}] & - & + \\[0.05em] \hline
& & & & & & \\[-1.4em]
[D_\mu, \hat{\chi}_+] & + & + & &
[D_\lambda,\langle F^+_{\mu\nu}\rangle ] & - & + \\[0.05em] \hline
\end{array}
\end{eqnarray*}
\caption{Building blocks of the relativistic pion--nucleon effective
theory. Note that the covariant derivative does not act on the nucleon
spinors not shown.}
\end{table}
Of course, hermiticity has to be assured by appropriate
factors of $i$ and combinations of terms. For the reasons given
above, we treat the covariant derivative acting on nucleons
separately and also the nucleon mass term, which is formally of
order $q^0$. Note, however, that the operator $i\barr{D}- m$ is 
proportional to the nucleon three--momentum, which can be small.

Now we have to construct the elements of the Clifford
algebra and all other operators of dimension zero/one to contract the 
Lorentz indices of the building blocks. Consider first all operators
constructed from $\gamma$ matrices, the metric tensor and the totally
antisymmetric tensor in $d=4$ according to the number of free
indices, called $N_I$:
\beqa\label{CNi}
N_0&:& \quad 1 \,\, ,\,\, \gamma_5 \,\, ; \quad N_1 : \quad
\gamma_\mu \,\, , \, \, \gamma_\mu \gamma_5 \,\, ; \quad N_2 : \quad
g_{\mu \nu} \,\, , \,\, g_{\mu \nu} \gamma_5  \,\, , \,\, 
\sigma_{\mu \nu} \,\, , \,\, \sigma_{\mu \nu} \gamma_5  \,\, ;
\nonumber \\
N_3&:& \quad  g_{\mu \nu} \gamma_\lambda \,\, , \,\, 
 g_{\mu \nu} \gamma_\lambda \gamma_5 \,\, , \,\, 
\epsilon_{\mu\nu\lambda\alpha} \gamma^\alpha \,\, , \,\,
\epsilon_{\mu\nu\lambda\alpha} \gamma^\alpha \gamma_5 \,\, .
\eeqa
Similarly, the terms involving the covariant derivative read
\beq\label{DNi}
N_1  : \,\,\,
D_\mu \,\, ;   \quad N_2 : \,\,\,
\{D_{\mu} , D_{\nu} \} \,\, ;
\quad N_3 : \,\,\,
D_{\mu} D_{\nu} D_\lambda \, {\rm + permutations} \,\,\, .
\eeq
It is important to note that $\gamma_5$ and $g_{\mu \nu} \gamma_5$
have chiral dimension one because they only connect the large to the
small components and thus appear first at order $1/m$.
Terms with more $D_\mu$'s can always be reduced to the ones listed
in eqs.(\ref{CNi},\ref{DNi}) or they only contribute to higher orders by
use of the baryon eom,
\beq\label{Neom}
i {\barr D} \Psi =\biggl( m  + {g_A \over 2} \gamma_5 \barr{u} \biggr) \,
\Psi + {\cal O}(q^2) \,\, .
\eeq
To be specific, consider as an example an operator $O$ of dimension three
constructed from the above building blocks and properly contracted
to be a scalar with $PC = ++$ contributing as $\bar{\Psi} \,  O \,
\Psi$ to the effective Lagrangian. As stated before, the covariant
derivative acting on a nucleon spinor is of order $q^0$, i.e. one could
have a term like $\bar{\Psi} \,  O \, D^2 \, \Psi +{\rm h.c.}$, 
which is also of dimension three
and commensurate with all symmetries. However, it contributes only to higher
orders as can be seen from the following chain of manipulations:
\beqa\label{Dex}
\bar{\Psi} \,  O \, D^2 \, \Psi &=& \bar{\Psi} \,  O \, 
g_{\mu\nu} D^\mu D^\nu \, \Psi \nonumber \\
&=&\bar{\Psi} \,  O \, \barr{D} \barr{D} \, \Psi -
i \, \bar{\Psi} \,  O \, \sigma^{\mu\nu} D_\mu D_\nu \, \Psi
  \nonumber \\
&=&\bar{\Psi} \,  O \, \barr{D} \barr{D} \, \Psi -
{i \over 2} \,\bar{\Psi} \,  O \, \sigma^{\mu\nu} [D_\mu , D_\nu] \,
  \Psi \,\, . 
\eeqa
Through a repeated use of the nucleon eom,
the first term on the r.h.s. of eq.(\ref{Dex}) amounts to 
three terms with coefficents $m^2$,
$m \,g_A$ and $g_A^2$, in order. While the first of these can 
simply be absorbed
in the LEC accompanying the original term $\bar{\Psi} O \Psi$, 
the other two start to
contribute at order four and five, respectively. Similarly, the second
term on the r.h.s. of  eq.(\ref{Dex}) leads to a dimension five operator
by use of eq.(\ref{Dcomm}). Equivalently, one can use
field transformations
to get rid off
these terms. This somewhat more elegant method is discussed in detail in
appendix~\ref{app:trafo}. 

It is now straightforward to combine the building
blocks with the operators to construct the
effective Lagrangian. One just has to multiply the various operators
so that the result has $J^{PC} = 0^{++}$. This gives
\beq \label{LpiN3r}
{\cal L}_{\pi N}^{(3)} = \sum_{i=1}^{51} \, \bar{\Psi} \, O_i^{(3)}
\, \Psi \quad ,
\eeq
where the $O_i^{(3)}$ are monomials in the fields of chiral 
dimension three.
The full list of terms is given in \cite{svenphd}. At this stage,
the Lagrangian is over--complete. All terms are allowed by the
symmetries, but they are not all independent. While one could work
with such an over--complete set, it is more economical to
reduce it to the minimal number of independent terms.

Let us now show how one can reduce the number of terms
in ${\cal L}_{\pi N}^{(3)}$. To be specific, consider the 
term $\bar\Psi \, \gamma_5 \, \gamma_\mu \, [i D_\mu , \chi_-] \, \Psi$.
Performing partial
integrations, we can reshuffle the covariant derivative to the 
extreme left and and right,
\beq
\bar\Psi \, \gamma_5 \, \gamma_\mu \, [i D_\mu , \chi_-] \, \Psi =
- \bar\Psi \, i \stackrel{\leftarrow}{D}_\mu \, \gamma_5 \, \gamma_\mu
\, \chi_- \, \Psi -  \bar\Psi \,  \gamma_5 \, \gamma_\mu
\, \chi_- \, i D_\mu \, \Psi \,\,\, .
\eeq
Using now the eom for $\Psi$, eq.(\ref{Neom}), and the one for $\bar{\Psi}$, this
can be cast into the form
\beq
\bar\Psi \, \gamma_5 \, \gamma_\mu \, [i D_\mu , \chi_-] \, \Psi =
- 2m \, \bar\Psi \, \gamma_5 \,
\, \chi_- \, \Psi + \frac{g_A}{2} \, \bar\Psi \,  [ \barr{u} ,
\chi_- ]  \, \Psi + {\cal O}(q^4) \,\,\, 
\eeq
and these are terms which already exist at dimension three, i.e.
the term under consideration can be absorbed in the structure
of these two terms. By use of the relations,\footnote{Here, the
symbol $\doteq$ means equal up to terms of higher order. For example,
if $A_1$ is an operator of dimension three, $A_1 \doteq
A_2$ stands for $A_1 = A_2 + {\cal O}(q^n),\,n\ge 4$.} 
\begin{eqnarray} 
\bar{\Psi} A^\mu i\,D_\mu \Psi + {\rm h.c.} & \doteq & 
2 m\,\bar{\Psi} \gamma_\mu A^\mu \Psi~,
\label{R1}
\\[0.3em]
\bar{\Psi} A^{\mu\nu} D_\nu D_\mu \Psi + {\rm h.c.} & \doteq &
-i\,m\left(\bar{\Psi} \gamma_\mu A^{\mu\nu} D_\nu \Psi + {\rm h.c.} \right)~,
\label{R2}
\\[0.3em]
\bar{\Psi} A^{\mu\nu\lambda} i\,D_\lambda D_\nu D_\mu \Psi + {\rm h.c.} 
& \doteq &
m\left(\bar{\Psi} \gamma_\mu A^{\mu\nu\lambda} D_\lambda D_\nu \Psi + 
{\rm h.c.} \right)~,
\label{R3}
\\[0.3em]
\bar{\Psi} \gamma_5 \gamma_\lambda A^{\mu\lambda} D_\mu \Psi + {\rm h.c.} 
& \doteq &
2m \bar{\Psi} \gamma_5 \sigma_{\mu\lambda} A^{\mu\lambda} \Psi 
+ \left(\bar{\Psi} \gamma_5 \gamma_\mu A^{\mu\lambda} D_\lambda \Psi 
+ {\rm h.c.} \right)~,
\label{R7}
\\[0.3em]
\bar{\Psi} \gamma_5 \gamma_\lambda A^{\mu\lambda\alpha} D_\alpha D_\mu \Psi +
{\rm h.c.}
& \doteq &
m\left(\bar{\Psi} \gamma_5 \sigma_{\mu\lambda} A^{\mu\lambda\alpha} D_\alpha \Psi + {\rm h.c.} \right)
\nonumber\\ & + & 
\left(\bar{\Psi} \gamma_5 \gamma_\mu A^{\mu\lambda\alpha} D_\alpha D_\lambda \Psi + {\rm h.c.} \right)~,
\label{R8}
\\[0.3em]
\bar{\Psi} \sigma_{\alpha\beta} A^{\alpha\beta\mu} i\,D_\mu \Psi + {\rm h.c.} 
& \doteq &
2m \bar{\Psi} \epsilon_{\alpha\beta\mu\nu} \gamma_5 \gamma^\nu A^{\alpha\beta\mu} \Psi
- \left(\bar{\Psi} \sigma_{\beta\mu} A^{\alpha\beta\mu}i\, D_\alpha \Psi + {\rm h.c.} \right)
\nonumber\\ & + &
\left(\bar{\Psi} \sigma_{\alpha\mu} A^{\alpha\beta\mu} i\,D_\beta \Psi + {\rm h.c.} \right)~,
\label{R9}
\\[0.3em] 
\bar{\Psi} \gamma_5 \sigma_{\alpha\beta} A^{\alpha\beta\mu} D_\mu \Psi + {\rm
  h.c.} 
& \doteq &
-2m \bar{\Psi} \gamma_5 \gamma^\beta A^{\mu\beta\mu} \Psi
-\left(\bar{\Psi} \gamma_5 \sigma_{\beta\mu} A^{\alpha\beta\mu} D_\alpha \Psi + {\rm h.c.} \right)
\nonumber\\ & + & 
2m \bar{\Psi} \gamma_5 \gamma^\alpha A^{\alpha\mu\mu} \Psi
+\left(\bar{\Psi} \gamma_5 \sigma_{\alpha\mu} A^{\alpha\beta\mu} D_\beta \Psi + {\rm h.c.} \right)~,
\label{R10}
\\[0.3em]
\bar{\Psi} \gamma_\mu [iD^\mu,A] \Psi & \doteq & 
\frac{g_A}{2} \bar{\Psi} \gamma^\mu \gamma_5 [A,u_\mu] \Psi~,
\label{PI1}
\\[0.3em]
\bar{\Psi} \gamma_5 \gamma_\mu [iD^\mu,A] \Psi & \doteq & 
-2m \bar{\Psi} \gamma_5 A \Psi  
-\frac{g_A}{2} \bar{\Psi} \gamma^\mu [A,u_\mu] \Psi~,
\label{PI2}
\end{eqnarray}
one can reduce the 51 terms down to 23 independent ones. The form of
the Lagrangian with independent terms only is called the ``minimal'' one.
This minimal Lagrangian is given by:
\beqa\label{LpiNrel} 
 {\cal L}_{\pi N}^{{\rm min}} &=& {\cal L}_{\pi N}^{(1)} +
 {\cal L}_{\pi N}^{(2,{\rm min})} + {\cal L}_{\pi N}^{(3,{\rm min})}~, 
\nonumber \\
{\cal L}_{\pi N}^{(2, {\rm min})} 
& = & \bar{\Psi} \,\, \bigg\{
c_1\,\langle \chi_+ \rangle
-\frac{c_2}{8m^2} \Big(\langle u_\mu u_\nu \rangle \{D^\mu,D^\nu\} +\mbox{h.c}
                  \Big)
+ \frac{c_3}{2} \langle u^2 \rangle 
\nonumber \\ &  & 
+ \frac{i\,c_4}{4} \sigma^{\mu\nu} [u_\mu,u_\nu]
+ c_5\,\hat{\chi}_+
+ \frac{c_6}{8m} \sigma^{\mu\nu} F^+_{\mu\nu}
+ \frac{c_7}{8m} \sigma^{\mu\nu} \langle F^+_{\mu\nu} \rangle
\bigg\} \, \Psi\qquad~, \nonumber \\
{\cal L}_{\pi N}^{(3, {\rm min})} 
&= & \bar{\Psi}  
\bigg\{ - \frac{d_1}{2m} \Big( [u_\mu,[D_\nu,u^\mu]] D^\nu 
                       + \mbox{h.c.} 
                 \Big)
- \frac{d_2}{2m} \Big( [u_\mu,[D^\mu,u_\nu]] D^\nu 
                       + \mbox{h.c.} 
                 \Big)
\nonumber \\ & &
+\,\frac{d_3}{12m^3} \Big( [u_\mu,[D_\nu,u_\lambda]] 
                           (D^\mu D^\nu D^\lambda + \mbox{sym.})
                           + \mbox{h.c.} 
                     \Big)
\nonumber \\ & &
-\,\frac{d_4}{2m} \Big( \epsilon^{\mu\nu\alpha\beta} 
                        \langle u_\mu u_\nu u_\alpha \rangle D_\beta
                        + \mbox{h.c.} 
                  \Big)
+ \frac{i\,d_5}{2m} \Big( [\chi_-,u_\mu] D^\mu
                          + \mbox{h.c.} 
                    \Big)
\nonumber \\ & &
+\,\frac{i\,d_6}{2m} \Big( [D^\mu,\hat{F}^+_{\mu\nu}] D^\nu
                           + \mbox{h.c.} 
                     \Big)
+ \frac{i\,d_7}{2m} \Big( [D^\mu,\langle F^+_{\mu\nu}\rangle] D^\nu 
                          + \mbox{h.c.} 
                    \Big)
\nonumber \\ & &
+\,\frac{i\,d_8}{2m} \Big( \epsilon^{\mu\nu\alpha\beta} 
                           \langle \hat{F}^+_{\mu\nu} u_\alpha \rangle D_\beta
                           + \mbox{h.c.} 
                     \Big)
+ \frac{i\,d_9}{2m} \Big( \epsilon^{\mu\nu\alpha\beta} 
                          \langle F^+_{\mu\nu} \rangle u_\alpha D_\beta
                          + \mbox{h.c.} 
                    \Big)
\nonumber \\ & &
+\,\frac{d_{10}}{2} \gamma^\mu \gamma_5 \langle u^2 \rangle u_\mu
+ \frac{d_{11}}{2} \gamma^\mu \gamma_5 \langle u_\mu u_\nu \rangle u^\nu 
- \frac{d_{12}}{8m^2} \Big( \gamma^\mu \gamma_5 \langle u_\nu u_\lambda \rangle
                            u_\mu \{D^\nu,D^\lambda\}
                            + \mbox{h.c.} 
                      \Big)
\nonumber \\ & &
-\,\frac{d_{13}}{8m^2} \Big( \gamma^\mu \gamma_5 \langle u_\mu u_\nu \rangle
                             u_\lambda \{D^\nu,D^\lambda\}
                             + \mbox{h.c.} 
                       \Big)
+ \frac{i\,d_{14}}{4m} \Big( \sigma^{\mu\nu}
                             \langle [D_\lambda,u_\mu] u_\nu \rangle D^\lambda
                             + \mbox{h.c.} 
                       \Big)
\nonumber \\ & &
+\,\frac{i\,d_{15}}{4m} \Big( \sigma^{\mu\nu}
                              \langle u_\mu [D_\nu,u_\lambda] \rangle D^\lambda
                              + \mbox{h.c.} 
                        \Big)
+ \frac{d_{16}}{2} \gamma^\mu \gamma_5 \langle \chi_+ \rangle u_\mu
+ \frac{d_{17}}{2} \gamma^\mu \gamma_5 \langle \chi_+ u_\mu \rangle 
\nonumber \\ & &
+\,\frac{i\,d_{18}}{2} \gamma^\mu \gamma_5 [D_\mu,\chi_-]
+ \frac{i\,d_{19}}{2} \gamma^\mu \gamma_5 [D_\mu,\langle \chi_- \rangle]
\nonumber \\ & &
-\,\frac{i\,d_{20}}{8m^2} \Big( \gamma^\mu \gamma_5 [\hat{F}^+_{\mu\nu},u_\lambda]
                                \{D^\nu,D^\lambda\}
                                + \mbox{h.c.} 
                          \Big)
+ \frac{i\,d_{21}}{2} \gamma^\mu \gamma_5 [\hat{F}^+_{\mu\nu},u^\nu]
\nonumber \\ & &
+\,\frac{d_{22}}{2} \gamma^\mu \gamma_5 [D^\nu,F^-_{\mu\nu}]
+ \frac{d_{23}}{2} \epsilon^{\mu\nu\alpha\beta} \gamma_\mu \gamma_5 
                   \langle u_\nu F^-_{\alpha\beta} \rangle
\bigg\}
\,\Psi\qquad~.
\eeqa
where we have also shown the terms of dimension two since the form
given in \cite{gss} is not minimal and can be reduced to the one shown here. 
We do not discuss the renormalization of this effective field theory
since the power counting can not be set up in a straightforward
way. A thorough and detailed discussion is given in ref.~\cite{gss}.

\subsection{Effective Lagrangian in the heavy fermion limit}

We are now in the position to perform the extreme non-relativistic
limit, i.e. to construct the heavy baryon effective field theory.
Our procedure follows closely the one in \cite{bkkm}, therefore
we only give some steps for completeness. Basically, one 
considers the mass of the nucleon large compared to
the typical external momenta transferred by pions or photons and writes
the nucleon four--momentum as
\beq p_\mu = m \, v_\mu + \ell_\mu \, , \quad p^2 = m^2 \, , \quad v \cdot
\ell \ll m \, . \eeq
Notice that later we have to differentiate between $m$
(the physical nucleon mass) and $\krig{m}$ (the nucleon mass in the
chiral limit) when we consider the nucleon
mass shift and wave function renormalization. 
Here, $v_\mu$ is the nucleon four--velocity (in the
rest--frame, we have $v_\mu =( 1 , \vec 0 \, )$). In that
case, we can decompose the wavefunction $\Psi (x)$ into velocity
eigenstates \cite{jm} \cite{bkkm} 
\beq \Psi (x) = \exp [ -i m v \cdot x ] \, [ H(x) + h(x) ] \eeq
with 
\beq \barr v \, H = H \,\, ,  \barr  v \, h = -h \,\, , \eeq
or in terms of velocity projection operators
\beq 
P_v^+ H = H \, , \, P_v^- h = h \, , \quad P_v^\pm =
\frac{1}{2}(1 \pm \barr v \,) \, , \quad P_v^+ + P_v^- = 1\,  . 
\eeq
One now eliminates the 'small' component $h(x)$ either by using the
equations of motion or path--integral methods.
The Dirac equation for the velocity--dependent
baryon field $H = H_v$ (we will always suppress the label '$v$') 
takes the form $i v \cdot \partial H_v = 0$ to lowest
order in $1/m$. This allows for a consistent chiral counting as described
and the effective pion--nucleon Lagrangian takes the
form:\footnote{Note that the parameters appearing in the effective
  Lagrangian should be taken at their values in the chiral limit.}  
\beq {\cal L}_{\pi N}^{(1)}  = \bar{H}
 \left( i v \cdot D  + {g}_A S \cdot u \right) H 
+ {\cal O}\left(\frac{1}{m} \right) \, , \label{lagr} 
\eeq
with $S_\mu$ the covariant spin--operator
\beq S_\mu = \frac{i}{2} \gamma_5 \sigma_{\mu \nu} v^\nu \, , \, 
S \cdot v = 0 \, , \, \lbrace S_\mu , S_\nu \rbrace = \frac{1}{2} \left(
v_\mu v_\nu - g_{\mu \nu} \right) \, , \, [S_\mu , S_\nu] = i
\epsilon_{\mu \nu \gamma \delta} v^\gamma S^\delta \,
 \, , \label{spin}\eeq
in the convention $\epsilon^{0123} = -1$. There is one subtlety to be
discussed here. In the calculation of loop graphs, divergences appear
which need to be regularized and renormalized. This is  most
easily done in dimensional regularization since it naturally preserves the
underlying symmetries. However, the totally antisymmetric Levi-Civita
tensor is ill-defined in $d \neq 4$ space--time dimensions. One
therefore has to be careful with the spin algebra. In essence,
one has to give a prescription to uniquely fix the finite pieces.
The mostly used convention is to only use the
anticommutator to simplify products of spin matrices and solely
take into account that the commutator is antisymmetric under
interchange of the indices. Furthermore, $S^2$ can be uniquely
extended to $d$ dimensions via $S^2 = (1-d)/4$. With that in mind,
two important observations can be made: Eq.(\ref{lagr}) does not
contain the nucleon mass term any more and also, all Dirac matrices
can be expressed as combinations of $v_\mu$ and $S_\mu$. With these
rules, which are given for completeness in appendix~\ref{app:cliff}, we can
translate all terms of the relativistic Lagrangian,
Eq.(\ref{LpiNrel}), into their heavy fermion counterpieces.
These terms are accompanied by LECs, which we call $d_i$, $i=1,\ldots,23$.

In addition, there are $1/m$ corrections to the various operators. 
These can be worked out  along the lines spelled out in
appendix~A of ref.\cite{bkkm}. As an example, consider the
part of the third order action which has the form
\beq
S_{\pi N}' = -{1 \over (2m)^2} \, \int d^4x \, \bar{H} \,( \gamma_0
\, {\cal B}^{(1)\dagger} \, \gamma_0 ) (i v \cdot D + g_A S \cdot u
) \, {\cal B}^{(1)} \, H \,\,\, ,
\eeq
which translates into the following piece of the
third order effective Lagrangian,
\beq
{\cal L}_{\pi N}^{(3)'}  = -{1 \over 4m^2} \bar{H} \,\biggl[ (i
\barr{D}^\perp + {1\over 2} g_A v \cdot u \gamma_5)(i v \cdot D + g_A
S\cdot u)( i \barr{D}^\perp - {1\over 2} g_A v \cdot u \gamma_5)
\biggr] H \,\, ,
\eeq
with $D^\perp_\mu = (g_{\mu\nu} -v_\mu v_\nu ) D^\nu$.
One now has to express the various products of derivatives, $\gamma$
matrices and so on, in terms of non--relativistic four--vectors.
For the product of the first terms in the three square brackets,
this amounts to 
\beq
\bar{H}  \barr{D}^\perp  v\cdot D  \barr{D}^\perp  H
= \bar{H} \bigl[ (v \cdot D)^3 + D_\mu  v\cdot D  D^\mu - 
2[S_\mu , S_\nu]\, D^\mu  v\cdot D  D^\nu \bigr ] H \,\, ,
\eeq
so that
\beq
{\cal L}_{\pi N}^{(3)'} = -{i \over 4m^2} \bar{H} \,\biggl[ (v \cdot D)^3
- D_\mu v\cdot D D^\mu + 2[S_\mu , S_\nu] D^\mu  v\cdot D  D^\nu 
+ \ldots \biggr] H \,\, .
\eeq
With 8 other relations of this type one can fill in the
ellipsis. Similarly, the other pieces contributing at this order
can be worked out. 

Finally, one has to renormalize the one--loop generating functional.
This can be done along the lines described in \cite{ecker} and
detailed in \cite{mm}. All divergences can be absorbed in the LECs
of the 23 terms previously constructed. In fact, only 14 of these
are divergent. For these, we introduce scale--dependent renormalized
values,
\beq
d_i = d_i^r (\lambda ) + {\kappa_i \over (4\pi F)^2} \, {\rm
  L}(\lambda ) \,\,  ,
\eeq
with
\beq
{\rm L} (\lambda ) = {\lambda^{d-4}\over 16\pi^2} 
\biggl\{ {1 \over d-4} - {1\over 2}
 \biggl[ \ln(4\pi ) + \Gamma '(1) +1 \biggr] \biggr\}  
\quad ,
\eeq
and $\lambda$ is the scale of dimensional regularization. In addition,
there are eigth terms which are only necessary for the
renormalization, i.e. that do not appear in matrix elements of
physical processes. To be precise, these terms stem solely from the
divergent part of the one--loop generating functional and have no
(finite) counterparts in the relativistic theory. Note also that all these terms
are proportional to the nucleon eom, $v \cdot D \,H$, and thus can be
transformed away by appropriate field redefinitions.

Putting pieces together, the complete third order 
Lagrangian takes the form
\beqa \label{lpin3f}
{\cal L}_{\pi N}^{(3)} &=& {\cal L}_{\pi N}^{(3),\, {\rm fixed}} +
\sum_{i=1}^{23} d_i \, \bar{H} \tilde{O}_i H 
+ \sum_{i=24}^{31} \tilde{d}_i \, \bar{H} \tilde{O}_i^{\rm div} H \,\,\, ,
\nonumber \\
& = & \bar{H} \left\{ {\cal O}^{(3)}_{\rm fixed} + {\cal O}^{(3)}_{\rm
    ct} + {\cal  O}^{(3)}_{\rm div}\right\} H \,\,\, ,
\eeqa
with
\begin{eqnarray}
{\cal O}^{(3)}_{\rm fixed} 
& = &
\frac{g_A}{8m^2}\,[D^\mu,[D_\mu,S\!\cdot\! u]]
-\frac{g_A^2}{32m^2}\,\epsilon^{\mu\nu\alpha\beta}v_\alpha S_\beta
\langle F^-_{\mu\nu} v\!\cdot\!u\rangle
-i\,\frac{1}{4m^2}\left(v\!\cdot\! D\right)^3
\nonumber \\[0.5em]
& - &
\frac{g_A}{4m^2}\,v\cdot\!\!\stackrel{\leftarrow}{D} S\!\cdot\!u\,v\!\cdot\! D
+\frac{1}{8m^2}\left(i\,D^2\, v\!\cdot\! D + \mbox{h.c.}\right)
\nonumber \\[0.5em]
& - &
\frac{g_A}{4m^2}\left(\{S\!\cdot\! D,v\!\cdot\! u\}\,v\!\cdot\! D + \mbox{h.c.}\right)
+\frac{3g_A^2}{64m^2}\left(i\langle(v\!\cdot\!u)^2\rangle\,v\!\cdot\!D + \mbox{h.c.}\right)
\nonumber \\[0.5em]
& + &
\frac{1}{32m^2}\left(\epsilon^{\mu\nu\alpha\beta}v_\alpha S_\beta
[u_\mu,u_\nu]\,v\!\cdot\! D + \mbox{h.c.}\right)
\nonumber \\[0.5em]
& - & 
\frac{1}{16m^2}\left(i\,\epsilon^{\mu\nu\alpha\beta}v_\alpha S_\beta
\hat{F}^+_{\mu\nu} v\!\cdot\! D + \mbox{h.c.}\right)
\nonumber \\[0.5em]
& - & 
\frac{g_A}{8m^2}\left(S\!\cdot\! u\,D^2 + \mbox{h.c.}\right)
-\frac{g_A}{4m^2}\left(S\cdot\!\!\stackrel{\leftarrow}{D} u\!\cdot\! D + \mbox{h.c.}\right)
\nonumber \\[0.5em]
& - &
\frac{1+2c_6}{8m^2}\,\left(i\,\epsilon^{\mu\nu\alpha\beta}v_\alpha S_\beta 
\hat{F}^+_{\mu\sigma}v^\sigma D_\nu + \mbox{h.c.}\right)
\nonumber \\[0.5em]
& - &
\frac{c_7}{4m^2}\left(i\,\epsilon^{\mu\nu\alpha\beta}v_\alpha S_\beta 
\langle F^+_{\mu\sigma}\rangle\,v^\sigma D_\nu + \mbox{h.c.}\right)
\nonumber \\[0.5em]
& + &
\frac{1+g_A^2+8m\,c_4}{16m^2}\left(\epsilon^{\mu\nu\alpha\beta}v_\alpha S_\beta
[u_\mu,v\!\cdot\! u] D_\nu + \mbox{h.c.}\right)
\nonumber \\[0.5em]
& + &
\frac{g_A}{32m^2}\left(i\,\epsilon^{\mu\nu\alpha\beta} v_\alpha
F^-_{\mu\nu} D_\beta + \mbox{h.c.}\right)
-\frac{g_A^2}{16m^2}\left(i\,v\!\cdot\!u\,u\!\cdot\!D  + \mbox{h.c.}\right)
\nonumber \\[0.5em]
& + &
i\,\frac{1+8m\,c_4}{32m^2}\,[v\!\cdot\! u,[D^\mu,u_\mu]]
+\frac{c_2}{2m}\left(i\langle v\!\cdot\!u\,u_\mu\rangle D^\mu + \mbox{h.c.}\right)~,
\end{eqnarray}

\begin{eqnarray}
{\cal O}^{(3)}_{\rm ct} 
& = &
i\,d_1(\lambda)\,[u_\mu,[v\!\cdot\! D,u^\mu]]
+ i\left(d_2(\lambda)-\frac{1+8m\,c_4}{32m^2}\right)
[u_\mu,[D^\mu,v\!\cdot\! u]]
\nonumber\\[0.5em]
& + &
i\left(d_3(\lambda)+\frac{g_A^2}{32m^2}\right)
[v\!\cdot\! u,[v\!\cdot\! D,v\!\cdot\! u]]  
\nonumber\\[0.5em]
& + &
i\left(d_4(\lambda)-\frac{g_A}{64m^2}\right)
\epsilon^{\mu\nu\alpha\beta}v_\alpha\langle u_\mu u_\nu
u_\beta\rangle
+ d_5(\lambda)\,[\chi_-,v\!\cdot\! u]
\nonumber\\[0.5em]
& + &
\left(d_6(\lambda)-\frac{1+2c_6}{16m^2}\right)
[D^\mu,\hat{F}^+_{\mu\nu}]\,v^\nu
+ \left(d_7-\frac{c_7}{8m^2}\right)
[D^\mu,\langle F^+_{\mu\nu}\rangle]\,v^\nu
\nonumber\\[0.5em]
& + &
\left(d_8(\lambda)+\frac{g_A}{64m^2}\right)
\epsilon^{\mu\nu\alpha\beta}v_\beta\langle \hat{F}^+_{\mu\nu} u_\alpha\rangle
\nonumber\\[0.5em]
& + &
\left(d_9+\frac{g_A}{64m^2}\right)
\epsilon^{\mu\nu\alpha\beta}v_\beta\langle F^+_{\mu\nu}\rangle\,u_\alpha
+ d_{10}(\lambda)\,S\!\cdot\! u\,\langle u\!\cdot\! u \rangle
\nonumber \\[0.5em]
& + &
d_{11}(\lambda)\,S^\mu u^\nu \langle u_\mu u_\nu\rangle
\nonumber\\[0.5em]
& + &
\left(d_{12}(\lambda)-\frac{4g_A(1+4m\,c_4)+g_A^3}{32m^2}\right)
S\!\cdot\! u\,\langle (v\!\cdot\! u)^2 \rangle
\nonumber\\[0.5em]
& + &
\left(d_{13}(\lambda)+\frac{2g_A(1+4m\,c_4)+g_A^3}{16m^2}\right)
S^\mu v\!\cdot\! u\,\langle u_\mu v\!\cdot\! u\rangle
\nonumber \\[0.5em]
& + &
d_{14}(\lambda)\,\epsilon^{\mu\nu\alpha\beta}v_\alpha S_\beta \langle[v\!\cdot\!
D,u_\mu]\,u_\nu \rangle
\nonumber\\[0.5em]
& + &
\left(d_{15}+\frac{g_A^2}{16m^2}\right)
\epsilon^{\mu\nu\alpha\beta}v_\alpha S_\beta \langle u_\mu 
[D_\nu,v\!\cdot\! u]\rangle
\nonumber \\[0.5em]
& + &
d_{16}(\lambda)\,S\!\cdot\! u\,\langle\chi_+\rangle
+ d_{17}\,S^\mu\langle u_\mu\chi_+\rangle
+ i\,d_{18}\,S^\mu[D_\mu,\chi_-]
\nonumber\\[0.5em]
& + &
i\,d_{19}\,S^\mu[D_\mu,\langle\chi_-\rangle]
+ i\left(d_{20}(\lambda)+\frac{g_A(1-c_6)}{8m^2}\right)
S^\mu v^\nu[\hat{F}^+_{\mu\nu},v\!\cdot\! u]
\nonumber\\[0.5em]
& + &
i\,d_{21}\,S^\mu [\hat{F}^+_{\mu\nu},u^\nu]
+ d_{22}\,S^\mu[D^\nu,F^-_{\mu\nu}]
+ d_{23}\,\epsilon^{\mu\nu\alpha\beta}S_\mu\langle u_\nu F^-_{\alpha\beta}\rangle~,
\end{eqnarray}

\begin{eqnarray}
{\cal O}^{(3)}_{\rm div} 
& = &
\tilde{d}_{24}(\lambda)\,i\left(v\!\cdot\!D\right)^3
+ \tilde{d}_{25}(\lambda)\,v\cdot\!\!\stackrel{\leftarrow}{D}
S\!\cdot\!u\,v\!\cdot\! D
+ \tilde{d}_{26}(\lambda)\left(i\,\langle u\!\cdot\!u\rangle\,v\!\cdot\!D +
  \mbox{h.c.}\right)
\nonumber \\[0.5em]
& + &
\tilde{d}_{27}(\lambda)\left(i\,\langle (v\!\cdot\!u)^2\rangle\,v\!\cdot\!D +
  \mbox{h.c.}\right)
+ \tilde{d}_{28}(\lambda)\left(i\,\langle\chi_+\rangle\,v\!\cdot\!D +
  \mbox{h.c.}\right)
\nonumber \\[0.5em]
& + &
\tilde{d}_{29}(\lambda)\left(S^\mu[v\!\cdot\!D,u_\mu]\,v\!\cdot\!D +
  \mbox{h.c.}\right)
\nonumber \\[0.5em]
& + &
\tilde{d}_{30}(\lambda)\left(\epsilon^{\mu\nu\alpha\beta}v_\alpha S_\beta
[u_\mu,u_\nu] \,v\!\cdot\!D + \mbox{h.c.}\right)
\nonumber \\[0.5em]
& + &
\tilde{d}_{31}(\lambda)\left(\epsilon^{\mu\nu\alpha\beta}v_\alpha S_\beta
\hat{F}^+_{\mu\nu}\,v\!\cdot\!D + \mbox{h.c.}\right)~.
\end{eqnarray}
Note that for the counterterms, one can also introduce
scale--independent LECs $\bar{d}_i$ via
\beq
d_i(\lambda) = \bar{d}_i + \frac{\kappa_i}{(4\pi F)^2} 
\left((4\pi)^2\,{\rm L} - \log\frac{\lambda}{M}\right)~,
\eeq
and the LECs for the operators needed for the renormalization only have,
of course, no finite piece,
\beq
\tilde{d}_i(\lambda)  =  \frac{\tilde{\kappa}_i}{(4\pi F)^2} 
\left((4\pi)^2\,{\rm L} - \log\frac{\lambda}{M}\right)~.
\eeq
Note that in ref.\cite{eckmoj} this Lagrangian
was first given in another basis. In that paper, however, there is
one term too much (their operator $O_4$ has a fixed coefficient $c_2/m$).
The form of the effective pion--nucleon Lagrangian given here is most
convenient for direct comparison with the many calculations done in
the basis of the review~\cite{bkkm}. The novel Feynman rules from the
dimension three Lagrangian relevant for the problem of pion--nucleon
scattering are collected in app.~\ref{app:rules}. For an easier
comparison with the Lagrangian given in \cite{eckmoj}, we have ordered
the operators in the same way. This means that their $b_i$ can be
mapped on our $d_i$ for $i=1,2,3$ and their $b_{i+1}$ onto $d_i$ for
$i=4,\ldots,23$. Note that the $b_i$ also contain some finite pieces
which are separated out explicitely in our dimension three Lagrangian.
We refer the reader to \cite{svenphd} for a more detailed comparison.
The $\beta$ functions $\kappa_i$ and $\tilde{\kappa}_i$ can be inferred
from ref.\cite{ecker}. For completeness, we list them here for our
basis:
\beqa
\kappa_1 &=& g_A^4/8 \,\, , \,\, \kappa_2 = -(1+5g_A^2)/12 \,\, , \,\,
\kappa_3 = (4-g_A^4)/8 \,\, , \,\, \kappa_4 = -g_A^3(4+3g_A^2)/16 \,\, ,
\nonumber \\
\kappa_5 &=& (1+5g_A^2)/24 \,\, , \,\, \kappa_6 = -(1+5g_A^2)/6 \,\, , \,\,
\kappa_8 = g_A^3/4 \,\, , \,\, \kappa_{10} = g_A(4-g_A^4)/8 \,\, ,
\nonumber \\
\kappa_{11} &=& g_A(6-6g_A^2+g_A^4)/12 \,\, , \,\, \kappa_{12} 
= -g_A(8-g_A^4)/8 \,\, , \,\,
\kappa_{13} = -g_A^5/12 \,\, , \nonumber \\
\kappa_{14} &=& -g_A^4/4 \,\, , \,\,
\kappa_{16} = g_A(4-g_A^2)/8 \,\, , \,\, \kappa_{20} = g_A \,\, , \nonumber \\
\tilde{\kappa}_{24} &=& -3g_A^2 \,\, , \,\, 
\tilde{\kappa}_{25} = g_A^3 \,\, , \,\, 
\tilde{\kappa}_{26} = -3g_A^2(4+3g_A^2)/16 \,\, , \,\,
\tilde{\kappa}_{27} = (8+9g_A^4)/16 \,\, , \nonumber \\
\tilde{\kappa}_{28} &=& -9g_A^2/16 \,\, , \,\,
\tilde{\kappa}_{29} = g_A^3/3 \,\, , \,\,
\tilde{\kappa}_{30} = -g_A^2(4+g_A^2)/8 \,\, , \,\,
\tilde{\kappa}_{31} = g_A^2 \,\, . 
\eeqa
Having constructed the effective pion--nucleon Lagrangian to order $q^3$, 
we now turn to apply it to elastic pion--nucleon scattering. To account 
for isospin breaking, one also has to extend this Lagrangain to include
virtual photons. This has already been done in~\cite{ms} and we refer the
reader to that paper. For a systematic study of isospin violation in the
elastic and charge exchange channels, one first has to find out to what 
accuracy the low energy $\pi N$ phase shifts can be described in the
isospin symmetric framework. This is the question which will be addressed in
the remaining sections of this investigation.

\section{Pion--nucleon scattering}
\setcounter{equation}{0}

\subsection{Basic definitions}

In the following, we consider the $\pi N$-scattering amplitude. 
In the center-of-mass system (cms)  the amplitude for the process
$\pi^a(q_1) + N(p_1) \to \pi^b(q_2) + N(p_2)$ takes the
following form (in the isospin basis): 
\beqa 
T^{ba}_{\pi N} &=& 
\biggl(\frac{E+m}{2m}\biggr) \, \biggl\lbrace \delta^{ba} 
\Big[ g^+(\omega,t)+ i \vec
\sigma \cdot(\vec{q}_2\times \vec{q}_1\,) \, h^+(\omega,t) \Big]
\nonumber \\ && \qquad\quad
+i \, \epsilon^{bac}
\tau^c \Big[ g^-(\omega,t)+ i \vec \sigma \cdot(\vec{q}_2 \times \vec{q}_1\,) \,
h^-(\omega,t) \Big] \biggr\rbrace 
\eeqa
with $\omega = v\cdot q_1 = v\cdot q_2$ the pion cms energy, $t=(q_1-q_2)^2$ 
the invariant momentum transfer squared, $E_1 = E_2 \equiv E = ( \vec{q\,}^2 +
m^2)^{1/2}$ the nucleon energy and
\beq
\vec{q\,}_1^2 = \vec{q\,}_2^2 \equiv
\vec{q\,}^2 = {(s-M^2-m^2)^2 -4m^2M^2 \over 4s} \quad .
\eeq
Furthermore, $g^\pm(\omega,t)$ refers to the
isoscalar/isovector non-spin-flip amplitude and $h^\pm(\omega,t)$ to the
isoscalar/isovector spin-flip amplitude. This form is most suitable
for the HBCHPT calculation since it is already defined in a
two--component framework.\footnote{These amplitudes should not be
called ``non-relativistic''. They are, however, more suitable for our
discussion than the amplitudes $A^\pm (s,t) ,B^\pm (s,t)$ which arise from a
manifest Lorentz--invariant formulation of the $\pi N$ scattering
amplitude.}  A more detailed discussion of the
pertinent kinematical relations (also for the case of different
nucleon and pion masses in the in-- and out--going states) is 
given in app.~\ref{app:kinematics}.
We note that Moj\v zi\v s \cite{moj} has introduced two--component
amplitudes $\alpha^\pm$ and $\beta^\pm$ which are related to the ones
we use via\footnote{Note, however, that in~\cite{moj} the amplitudes
are evaluated in a particular frame, which is not the cms.
Therefore, care has to be taken when comparing our amplitudes
with the ones of Moj\v zi\v s.}
\beq
g^\pm = \alpha^\pm \,\, , \quad h^\pm = -\frac{1}{2} \beta^\pm \,\,\, .
\eeq
The difference in the spin--flip amplitude is due to our use of the
Pauli matrices whereas in  \cite{moj} the spin--vector $S_\mu$ was used.
The partial wave amplitudes $f_{l\pm}^\pm (s)$, where $l$ refers to the
orbital angular momentum, the superscript '$\pm$' to the isospin
and the subscript '$\pm$' to the total angular momentum ($j=l\pm s$), are given
in terms of the invariant amplitudes via
\beq
f_{l\pm}^\pm (s) = {E+m \over 16 \pi \sqrt{s}} \, \int_{-1}^{+1} dz \, \biggl[\,
g^\pm \, P_l (z) + \vec{q\,}^2 \, h^\pm \, (P_{l\pm1}(z) -zP_l (z) ) \biggr]~,
\eeq
where $z = \cos(\theta)$ is the angular variable and $\sqrt{s}$ the
total cms energy. 
The $P_l(z)$ are the conventional Legendre polynoms. For a given
isospin $I$, the phase shifts
$\delta_{l\pm}^I (s)$ can be extracted from the partial waves via
\beq
f_{l\pm}^I (s) = {1\over 2i | \vec{q\,}|} \, \biggl[ \exp(2i
\delta_{l\pm}^I) - 1 \biggr]~.
\eeq
For vanishing inelasticity, which is the case for the energy range
considered in this work,  the phase shifts are real. They are given
by (for a detailed discussion about the issue of defining the phase
shifts in an effective field theory, see ref.\cite{game})
\beq
\delta_{l\pm}^I (s) = \arctan \bigl( |\vec{q}\,|\, 
{\rm Re}~f_{l\pm}^I (s) \bigr)~.
\eeq
Equally well, one could use the definition without the arctan, the difference
being of higher order. For the phase shifts in the kinematical region considered
below, this difference is negligible. 

What we are after is the chiral expansion of the various amplitudes
$g^\pm, h^\pm$. These consist of essentially three pieces, which are
the Born and counterterm parts of polynomial type as well as the
unitarity corrections due to the pion loops. To be precise, we
have
\beq\label{chexp}
X = X^{\rm tree} + X^{\rm ct} + X^{\rm loop} \,\, , \quad
X= g^\pm  ,  h^\pm \,\, ,
\eeq
where the tree contribution subsumes all terms generated from the
lowest order Lagrangian, the counter term amplitude the ones
proportional to the dimension two and three LECs (i.e. $\sim c_i ,
c_i/m , d_i)$ and the last term in Eq.(\ref{chexp}) is the one--loop
order $q^3$ amplitude. The latter is a complex--valued function and
restores unitarity in the perturbative sense. Its various terms
are all proportional to $1/F^4$. 
These amplitudes are functions of two kinematical variables.
We choose the pion energy and the invariant
momentum transfer squared, i.e. $X= X (\omega, t)$. In what
follows, we mostly suppress these arguments.
Before discussing the
full order $q^3$ one--loop amplitude, we wish to elaborate on the
relativistic tree graphs with insertions from the dimension one
effective Lagrangian. These terms are all proportional to $g_A^2/F^2$
or $1/F^2$.

\subsection{Evaluation of the  tree graphs}

In this section we wish to elaborate on the tree amplitudes $X^{\rm
tree}$. It was already stated in \cite{bkmzpc} that the most
convenient way to calculate the chiral expansion of these is to take
the relativistic theory and expand the resulting amplitudes in powers
of $1/m$ to the order one is working. Ecker and Moj\v zi\v
s~\cite{eckmwf} have given a different prescription for calculating in
the heavy fermion formalism. Their procedure involves an unusual,
momentum--dependent wavefunction renormalization. We show here that
one can indeed recover the relativistic result from the heavy fermion
theory by straightforward calculation to third order and for the
process under consideration.  A general proof of the equivalence
between this method and  the one proposed in \cite{eckmwf} 
is given in ref.\cite{zfac}

Consider first the relativistic tree graphs shown in fig.~1. These are
the nucleon pole graph (fig.~1a), its crossed partner (fig.~1b) and
the seagull term (fig.~1c) involving the non--linear pion--nucleon 
$\bar{N}N\pi\pi$ vertex. Using the notation of
app.~\ref{app:kinematics}, their respective contributions are
\beqa
T^{(a)}_{\pi N} &=& \frac{g_A^2}{4F^2} \, \tau^b \tau^a \, \bar{u} (p_2)\,
\barr{q}_2 \, \frac{-\barr{p}_1 - \barr{q}_1 + m}{(p_1 +q_1)^2-m^2} \,
\barr{q}_1 \, u(p_1)~, \nonumber \\
T^{(b)}_{\pi N} &=& \frac{g_A^2}{4F^2} \, \tau^a \tau^b \, \bar{u} (p_2)\,
\barr{q}_1 \, \frac{-\barr{p}_1 + \barr{q}_2 + m}{(p_1 -q_2)^2-m^2} \,
\barr{q}_2 \, u(p_1)~, \nonumber \\
T^{(c)}_{\pi N} &=& -\frac{i}{4F^2} \,\epsilon^{abc} \tau^c\,  \bar{u} (p_2)\,
( \barr{q}_1 +  \barr{q}_2) \,  u(p_1)~.
\eeqa
Note that while the first two terms have an isoscalar and an isovector
contribution, the seagull term is purely isovector. It is also
important to remember that all quantities appearing in these formulas,
such as the nucleon mass, the axial-vector coupling etc., are taken at
their chiral limit values. For a more compact notation, we refrain
from denoting these by the symbol '$^\circ$'. The nucleon spinors are
normalized as given in Eq.(\ref{spinor}) and we will not expand
these normalization factors in what follows. Straightforward
algebra allows one to extract the amplitudes $g^\pm, h^\pm$ from these
expressions. Putting all pieces together and expanding
up--to--and--including order $1/m^2$, we find
\beqa \label{treeamp}
g^+ (\omega, t)  &=& -\frac{g_A^2}{F^2}
\frac{1}{16m\omega^2} \biggl[ 4M^4 + t^2 + 4 \omega^2 t - 4M^2 t \biggr]
\nonumber \\
&+& \!\!\!
\frac{g_A^2}{F^2} \frac{1}{32m^2\omega^3} \biggl[
16\omega^2 M^4 + 5\omega^2 t^2 - 16M^6 -8M^2t^2 + 4\omega^4 t + 20 M^4 t
\nonumber \\ &&\qquad\qquad\qquad\qquad\qquad\qquad
+ t^3 - 20 \omega^2 M^2 t \biggr]~,  \\  
g^- (\omega, t)  &=& -\frac{g_A^2}{F^2} \frac{1}{4\omega }\biggl[
2\omega^2 - 2M^2 + t \biggr] + \frac{\omega}{2F^2} \nonumber \\
&+& \!\!\! \frac{g_A^2}{F^2}  \frac{1}{16m\omega^2} \biggl[ t^2 + 
2\omega^2 t - 8\omega^4 + 8M^4 -6M^2 t \biggr] -\frac{1}{8mF^2}
\biggl[ 4M^2 - 4\omega^2 -t \biggr] \nonumber \\
&+& \!\!\!
\frac{g_A^2}{F^2} \frac{1}{32m^2\omega^3} \biggl[8M^2 \omega^4 - 20M^4
\omega^2 -4\omega^6 + 20 M^6 + 20 \omega^2 M^2 t - 5\omega^2 t^2
-4t \omega^4 \nonumber \\
&&  \qquad\qquad\quad
-22M^4 t + 8t^2M^2 -t^3 \biggr] + \frac{\omega}{16m^2F^2} \biggl[ 2 \omega^2 -
2M^2 + t \biggr]~, \\
h^+ (\omega, t)  &=& -\frac{g_A^2}{F^2} \frac{1}{2\omega} 
- \frac{g_A^2}{F^2} \frac{1}{8 m\omega^2} \biggl[ 4M^2-t \biggr]
\nonumber \\
&-& \!\!\! 
\frac{g_A^2}{F^2} \frac{1}{16 m^2\omega^3}
\biggl[10M^4+t^2-6\omega^2M^2 -2\omega^4 + 3t\omega^2-6M^2t \biggr]~, \\
\label{treeampend}
h^- (\omega, t)  &=& -\frac{g_A^2}{F^2} \frac{1}{8 m\omega^2} 
\biggl[ 2\omega^2 - 2M^2 +t \biggr] + \frac{1}{F^2}\frac{1}{4m}
\nonumber \\
&-& \!\!\! 
\frac{g_A^2}{F^2} \frac{1}{16 m^2\omega^3}
\biggl[2\omega^4 -t^2 -8M^4 +6M^2 t - 3\omega^2 t +6\omega^2 M^2 \biggr]
+\frac{\omega}{8m^2F^2} \,\, . 
\eeqa

Consider now the equivalent graphs in the heavy fermion language. They
are shown in fig.~2. The heavy dot (box) denotes a dimension two
(three) insertion with a {\it fixed} coefficient, i.e. all LECs are
set to zero (they contribute to $X^{\rm ct}$). Using the Feynman rules
of app.~\ref{app:rules}, one can work out these diagrams
straightforwardly. We stress that we take the same
normalization factor as for the relativistic calculation, 
which again is not expanded. This is different from the treatment in ref.~\cite{moj}.
We give here only some illustrative intermediate results.
The T--matrix corresponding to diagrams~1a,2a, and 3a reads:
\beqa
T_{\pi N}^{(1a)} &=& -{\cal N}^2\frac{g_A^2}{F^2} \, \tau^b \tau^a \,
S \cdot q_2 \,S\cdot q_1 \, \frac{1}{v \cdot (p_1+q_1)}~,\nonumber \\
T_{\pi N}^{(2a)} &=& {\cal N}^2\frac{g_A^2}{F^2} \, \frac{1}{2m}\, \tau^b \tau^a \,
 S \cdot q_2 \, S \cdot (2p_2 + 2q_2-q_1) \frac{v\cdot q_1}{v\cdot (p_1+q_1)}~,
\nonumber \\
T_{\pi N}^{(3a)} &=& {\cal N}^2\frac{g_A^2}{F^2} \, \frac{1}{8m^2}\, \tau^b \tau^a \,
\, \frac{1}{v \cdot (p_1+q_1)} \, S \cdot q_2 \,S\cdot q_1 \,
\nonumber\\
&\times& 2\,\Bigl[
v\cdot p_1 \, v\cdot (p_1+q_1) - (p_1 + q_1)^2
+v\cdot (2p_1+q_1) v\cdot q_1 \Bigr]~,
\eeqa
with ${\cal N}$ defined in Eq.(\ref{spinor}) (here for the equal mass case).
Dissecting these  into spin--dependent and independent terms,
we get (including also the crossed diagrams)
as contributions to the invariant amplitudes in the cms:
\beqa
g^\pm_{(1a)} &=& -\frac{g_A^2}{F^2} \frac{1}{8} (2\omega^2 - 2M^2 + t)
\, \biggl[ \frac{1}{v\cdot (p_1+q_1)} \pm \frac{1}{v\cdot (p_1-q_2)}
\biggr] \,\, ,  \\
h^\pm_{(1a)} &=& \frac{g_A^2}{F^2} \frac{1}{4} 
 \biggl[ \frac{-1}{v\cdot (p_1+q_1)} \pm \frac{1}{v\cdot (p_1-q_2)}
\biggr] \,\, , \\
g^\pm_{(2a)} &=& \frac{g_A^2}{F^2}\frac{1}{2m} \biggl[
 -\frac{1}{8} (2\omega^2 - 2M^2 + t)  \frac{v\cdot q_1}{v\cdot
   (p_1+q_1)} \nonumber \\
&& \qquad \quad\pm \biggl( 2 S\cdot q_1 S\cdot p_1-\frac{1}{8} 
(2\omega^2 - 2M^2 + t) \biggr)
\frac{v\cdot q_2}{v\cdot (p_1-q_2)}
\biggr] \,\, ,  \\
h^\pm_{(2a)} &=& -\frac{g_A^2}{F^2}\frac{1}{8m} \biggl[
 \frac{v\cdot q_1}{v\cdot  (p_1+q_1)} \mp
\frac{v\cdot q_2}{v\cdot (p_1-q_2)}
\biggr] \,\,  ,  \\
g^\pm_{(3a)} &=& \frac{g_A^2}{F^2}\frac{1}{8m^2}\biggl\lbrace
\frac{1}{8} (2\omega^2 - 2M^2 + t)  \biggl[ \frac{1}{v\cdot (p_1+q_1)}
(q_1^2 + 2v\cdot p_1 v\cdot (p_1+q_1) \nonumber \\
&-& (p_1+q_1)^2 - p_1^2
+2v \cdot(2p_1+q_1) v\cdot q_1 - 2q_1 \cdot (p_1+q_1) \bigr) \nonumber \\
&\pm& \frac{1}{v\cdot (p_1-q_2)}(q_2^2 + 2v\cdot p_1 v\cdot (p_1-q_2) 
- (p_1-q_2)^2 - p_1^2 \nonumber \\
&+& 2v \cdot(2p_1-q_2) v\cdot q_2 - 2q_2 \cdot p_1 \bigr) \biggr] \pm
\frac{1}{2}
\frac{\omega_1^2 - q_1^2}{v\cdot(p_1-q_2)} \times \nonumber \\
&&\qquad\qquad\qquad\qquad \biggl[  2v \cdot(2p_1-q_2) v\cdot q_2 -
q_2 \cdot (2p_1-q_2) \biggr]\biggr\rbrace~, \\
h^\pm_{(3a)} &=& \frac{g_A^2}{F^2}\frac{1}{32m^2}\biggl\lbrace
\frac{1}{v\cdot (p_1+q_1)}
(q_1^2 + 2v\cdot p_1 v\cdot (p_1+q_1) - (p_1+q_1)^2 - p_1^2\nonumber \\
&& \qquad\qquad\qquad
+ 2v \cdot(2p_1+q_1) v\cdot q_1 - 2q_1 \cdot (p_1+q_1)) \nonumber \\
&\mp& \frac{1}{v\cdot (p_1-q_2)}(q_2^2 + 2v\cdot p_1 v\cdot (p_1-q_2) 
- (p_1-q_2)^2 - p_1^2 \nonumber \\
&& \qquad\qquad\qquad\qquad\qquad
+ 2v \cdot(2p_1-q_2) v\cdot q_2 - 2q_2 \cdot p_1)  \biggr\rbrace~.
\eeqa
The other diagrams shown in fig.~2 can be evaluated similarly.
Putting all pieces together, one {\it exactly} recovers the 
amplitude calculated by expanding the relativistic tree graphs, cf. 
Eq.(\ref{treeamp}-\ref{treeampend}). 
We are presently investigating other observables
like the electromagnetic form factors and so on to show that this method
can be employed in any situation. This is indeed proven in ref.\cite{zfac}.

\subsection{One loop result for the amplitudes}

The full one--loop amplitude to order $q^3$ is obtained after mass
and coupling constant renormalization,
\beq
(\krig{g}_A , \krig{m}, F , M) \to ({g}_A , {m}, F_\pi , M_\pi )~.
\eeq
The pertinent formulas can be found in refs.~\cite{bkmrev,bkmpipin}.
The tree level amplitudes are
given by the expressions of Eqs.(\ref{treeamp}-\ref{treeampend}), with the important
difference that all parameters are taken at their {\it physical}
values, not the ones in the chiral limit. For the construction of the
loop and the counterterm amplitudes, we use the scale--independent
LECs $\bar{d}_i$, so that in our final expressions all terms $\sim
\ln( M_\pi /\lambda  )$ have disappeared. Nevertheless, our amplitude
agrees with the one given in appendix~A of \cite{bkmlec}. We have:
\beqa
g^+_{\rm ct} &=& \frac{1}{F^2}\Bigl[ -4c_1 M^2 + 2c_2 \omega^2
+ c_3(2M^2-t) \Bigr] + \frac{c_2 \, \omega}{m F^2} 
\Bigl[ 4 \omega^2 - 4M^2 +t \Bigr]~, \\
g^+_{\rm loop} &=& i\,\frac{\omega^2}{8 \pi F^4}\sqrt{\omega^2
  -M^2} + \frac{g_A^2}{F^4}\frac{1}{32 \pi} \Bigl( M^2 -2t \Bigr)
\Bigl(M+ \frac{2M^2-t}{2\sqrt{-t}}\arctan \frac{\sqrt{-t}}{2M}\Bigr)
\nonumber \\
&&\qquad\qquad + \frac{g_A^4}{F^4}\frac{1}{24 \pi \omega^2} \Bigl(
t-2M^2+2\omega^2\Bigr) \, \Bigr[ i (\omega^2-M^2)^{3/2} -M^3
\Bigr]~,\\
g^-_{\rm ct} &=& \frac{c_4 \, \omega \, t}{2m F^2} 
+ \frac{2\omega}{F^2} (2M^2-t) \, (\bar{d}_1+\bar{d}_2 )
+ \frac{4\omega^3}{F^2} \, \bar{d}_3 +\frac{8 \omega\, M^2}{F^2} \, \bar{d}_5
\nonumber \\ &&
\qquad\qquad 
+ \,
\frac{g_A}{F^2}\frac{M^2}{\omega} (2\omega^2-2M^2+t)\,
\bar{d}_{18}~, \\
g^-_{\rm loop} &=& \frac{1}{32 \pi^2 F^4} \biggl\{ \frac{\omega}{3}
(4M^2-t) \sqrt{1 - \frac{4M^2}{t}} \ln
\frac{\sqrt{4M^2-t}+\sqrt{-t}}{2M} \nonumber \\
&& \!\!\!\!\!\!\!\!\!\!\!\! 
-4\omega^2 \sqrt{\omega^2-M^2} \Bigl[-i\frac{\pi}{2} + 
\ln\Bigl( \frac{\omega}{M} + \sqrt{\frac{\omega^2}{M^2} -1}
\Bigr)\Bigr] + \frac{\omega}{9} \Bigl(18\omega^2 -12M^2 + \frac{5}{2}t
\Bigr)\biggr\}
\nonumber \\
&& \!\!\!\!\!\!\!\!\!\!\!\!
 + \frac{g_A^2}{24 \pi^2 F^4}\frac{1}{\omega} \biggl\{ -\frac{1}{4}
\omega^2 (5t -8M^2) \sqrt{1 - \frac{4M^2}{t}} \ln
\frac{\sqrt{4M^2-t}+\sqrt{-t}}{2M} - 2\omega^2 M^2+
\frac{13}{24}\omega^2 t \biggr\} \nonumber \\
&& + \frac{g_A^4}{96 \pi^2 F^4}\frac{1}{\omega^2}(2M^2-2\omega^2 -t)
\biggl\{ -2 (\omega^2-M^2)^{3/2} \Bigl[ i \frac{\pi}{2} -
\ln\Bigl( \frac{\omega}{M} + \sqrt{\frac{\omega^2}{M^2} -1}
\Bigr)\Bigr] \nonumber \\ && \qquad \qquad \qquad \qquad \qquad 
\qquad \qquad \qquad 
+ 2M^2 \omega - \frac{5}{3}\omega^3 \biggr\}~, \\
h^+_{\rm ct} &=& \frac{2 \omega}{F^2} \, (\bar{d}_{14} - \bar{d}_{15}
) + \frac{g_A}{F^2} \frac{2M^2}{\omega}\,\bar{d}_{18}~,\\
h^+_{\rm loop} &=& -\frac{g_A^4}{24 \pi^2 F^4}\frac{1}{\omega^2}
\biggl\{ M^2 \omega + \frac{\omega^3}{6} - 
(\omega^2-M^2)^{3/2}  \Bigl[ i \frac{\pi}{2} -
\ln\Bigl( \frac{\omega}{M} + \sqrt{\frac{\omega^2}{M^2} -1}
\Bigr)\Bigr] \biggr\}~, \\
h^-_{\rm ct} &=& \frac{c_4}{F^2} + \frac{c_4 \, \omega}{m F^2}~,\\
h^-_{\rm loop} &=& \frac{g_A^2}{32 \pi F^4} \Bigl[ -M +  
\frac{t -4M^2}{2\sqrt{-t}}\arctan \frac{\sqrt{-t}}{2M} 
+ \frac{4g_A^2}{3\omega^2} \Bigl( i (\omega^2-M^2)^{3/2} - M^3 \Bigr)
\Bigr]~.
\eeqa
One can easily convince oneself that the imaginary parts stemming from the
loop contributions at order $q^3$ fulfill perturbative unitarity. In terms of
the partial wave amplitudes this reads:
\beq
{\rm Im}~f_l^{I(3)} (s) = |\vec{q}_\pi \,| \biggl( {\rm Re}~f_l^{I(1)}
(s) \biggr)^2~.
\eeq
This is an important check of the calculation. The analytic structure
of these amplitudes is further discussed in app.~A of
ref.\cite{bkmlec}. Note also that the amplitudes have the proper
crossing behaviour under $\omega_{\rm lab} \to -\omega_{\rm lab}$.
In the formulas above, $\omega$ is given in the cms, but we remark
that to the order we are working, $\omega_{\rm lab} = \omega_{\rm cms}$. 
More precisely, $\omega_{\rm lab} = \omega_{\rm cms} + (\omega_{\rm
  cms}^2 - M^2)/m +\,$higher orders. 

\section{Results}
\setcounter{equation}{0}

\subsection{The fitting procedure}

Before presenting the actual results, we have to discuss how to fix the
four dimension two LECs and the five (combinations of) dimension three
LECs. One way would be to fit to threshold and subthreshold parameters.
Up to date, only the Karlsruhe (KA) analysis gives these consistently
with the phase shifts~\cite{hoeh}. 
From the point of the chiral expansion, this expansion
around the center of the Mandelstam triangle together with the threshold
parameters should provide the best data to
fix the LECs. Since, however, some of the data used in the KA analysis
have become doubtful (for a detailed discussion see e.g. ref.\cite{me97}), we 
prefer to fit directly to the available six S-- and P--wave
phase shifts in a certain energy
range, as discussed below. The threshold and subthreshold parameters are
then predicted. We remark that there still is debate about the actual value
of the isoscalar S--wave scattering length as extracted from 
pionic atoms~\cite{sigg}
or the various partial--wave amplitudes. 
As input we use the phase shifts of the 
Karlsruhe (KA85) group~\cite{koch}, from the analysis of
Matsinos~\cite{mats} (EM98)
and the latest update of the VPI group, called SP98~\cite{SAID}. Since the
Karlsruhe and the VPI groups give no errors for the phase shifts,
we have assigned a common uncertainty of $\pm 3$\% to all values. This procedure
is, of course, somewhat arbitrary. We have convinced ourselves that assigning
a different error to the phase shifts does not change the results but only the
respective $\chi^2$/dof. In case of the EM98 phases, we have used the given
errors which vary for the various channels. Again, we find that the results
are consistent with the ones for the other phases which gives us some
confidence that our procedure is not quite useless. Preferrably, we would
have worked with uncertainties provided directly by the various groups.
This issue deserves further study. Note also that the LEC $\bar{d}_{18}$
is fixed by means of the Goldberger--Treiman discrepancy, i.e.
by the value for the pion--nucleon coupling constant extracted in the various
analyses,
\beq
g_{\pi N}= {g_A \, m\over F_\pi} \biggl( 1 - {2M_\pi^2\, \bar{d}_{18}\over g_A}
\biggr) \,\,.
\eeq
The actual values of $g_{\pi N}$ are 
\beq
g_{\pi N} = 13.4\pm 0.1 \,, \,\,\, 13.18 \pm 0.12\, ,\,\,\,
13.13\pm 0.03 \, ,
\eeq
for KA85, EM98 and SP98. Throughout, we use $g_A = 1.26$, $F_\pi =
92.4\,$MeV, $m = 938.27\,$MeV and $M_\pi = 139.57\,$MeV. 
Finally, we remark that we
do not use the value of the pion--nucleon $\sigma$--term in the fitting procedure.
This would to a large extent determine the value of the LEC $c_1$. Due to the
present discussion about the empirical value of the 
pion--nucleon $\sigma$--term~\cite{me97},
we prefer not to use it in the fit. We believe, however, that the analysis
of \cite{gls} still gives the most trustworthy value for $\sigma_{\pi N}(0)$.

\subsection{Phase shifts and threshold parameters}

After the remarks of the preceeding paragraph, we can now present results.
For the KA85 case, we have fitted the data up to 100~MeV pion  lab
momentum (i.e. 4 points per partial wave at $q_\pi = 40, 60, 79,
97$~MeV). For the analysis
of Matsinos, we use 17 points for each partial wave in the range
of $q_\pi = 41.4 - 96.3\,$~MeV 
for the fitting. For the VPI SP98 solution, one has all phases down to
threshold. However, as pointed out to us by Pavan, the values at lower values
of $q_\pi$ are somewhat uncertain. Indeed, our fits underline this statement.
Using only the values between threshold and $q_\pi=50$~MeV, we are not able to
get a low-$\chi^2$ fit. If we use the 5 data points in the range between 60 and
100~MeV  we get a stable fit. We also checked that extending the fitting range
to higher energies, the fits become unstable due to the Delta
resonance.  From here on, we call the fits corresponding to the
Karlsruhe, Matsinos and VPI analysis, fit~1, 2 and 3, in order. 
The resulting LECs  are given in table~\ref{tab:LEC}. 
\renewcommand{\arraystretch}{1.2}
\begin{table}[h]
\begin{center}
\begin{tabular}{|c|c|c|c|}
    \hline
    LEC      &  Fit 1  &  Fit 2  & Fit 3  \\
    \hline\hline
$c_1$ & $-1.23\pm 0.16 $ & $-1.42\pm 0.03 $ & $-1.53\pm 0.18$ \\
$c_2$ & $ 3.28\pm 0.23 $ & $ 3.13\pm 0.04 $ & $ 3.22\pm 0.25$ \\
$c_3$ & $-5.94\pm 0.09 $ & $-5.85\pm 0.01 $ & $-6.20\pm 0.09$ \\
$c_4$ & $ 3.47\pm 0.05 $ & $ 3.50\pm 0.01 $ & $ 3.51\pm 0.04$ \\
\hline
$\bar{d}_1+\bar{d}_2$ 
      & $ 3.06\pm 0.21 $ & $ 3.31\pm 0.14 $ & $ 2.68\pm 0.15$ \\
$\bar{d}_3$ 
      & $-3.27\pm 0.73 $ & $-2.75\pm 0.18 $ & $-3.11\pm 0.79$ \\
$\bar{d}_5$ 
      & $ 0.45\pm 0.42 $ & $-0.48\pm 0.06 $ & $ 0.43\pm 0.49$ \\
$\bar{d}_{14}-\bar{d}_{15}$ 
      & $-5.65\pm 0.41 $ & $-5.69\pm 0.28 $ & $-5.74\pm 0.29$ \\
$\bar{d}_{18}$ 
      & $-1.40\pm 0.24 $ & $-0.78\pm 0.27 $ & $-0.83\pm 0.06$ \\
   \hline\hline
  \end{tabular}
  \caption{Values of the LECs in GeV$^{-1}$ and GeV$^{-2}$ for 
           the $c_i$ and $\bar{d}_i$, respectively, for the various fits
           as described in the text.\label{tab:LEC}}
\end{center}\end{table}
Note that the error on the LECs is purely 
the one given by the fitting routine and is certainly underestimated.
All the LECs come out of ``natural size'', i.e. they are numbers
of order one. For that, one introduces the dimensionless LECs
$c_i' = 2m c_i$ and $\bar{d}_i\, ' = 4m^2 \bar{d}_i$. Inspection of the
values given above shows that all LECs turn out to be between one
and ten (up to a sign, of course). A more detailed discussion of the
values of these LECs is given in the next paragraph. 
The resulting S-- and P--wave phase shifts are shown in figs.~3~(fit~1),
4~(fit~2) and 5~(fit~3). The corresponding $\chi^2$/dof is 0.83,
0.77 and 1.34 for fits~1,2 and 3, respectively. Note, however, that
only in the case of fit~2 this number is really meaningful since it
reflects the uncertainties of the data as given in the analysis.
The description of the phase shifts is surprisingly good even at
higher energies and also for the P--waves. Even  the bending of
the $P_{11}$ phase due to the Roper resonance can be
reproduced in fit~3 but is clearly absent in fit~2.
Furthermore, the tail of the $\Delta$ in the $P_{33}$ channel is
underestimated. Still, in an overall view these six lowest partial
waves can be described fairly well in this ${\cal O}(q^3)$
calculation up to surprisingly large pion momenta. 

\begin{table}[hbt]
\begin{center}
\begin{tabular}{|c||c|c|c||c|c|c|}
    \hline
   Obs.    &  Fit 1  &  Fit 2  & Fit 3  & KA85 & EM98 & SP98 \\
    \hline\hline
$a^+_{0+}$  & $-0.97$ & $0.49$ & $0.25$  
      & $-0.83$ & $ 0.41\pm 0.09 $ & $ 0.0 \pm 0.1$ \\
$b^+_{0+}$  & $-4.77$ & $-5.23$ & $-6.33$  
      & $-4.40$ & $ -4.46  $ & $-4.83 \pm 0.10$ \\
$a^-_{0+}$  & $9.05$ & $7.72$ & $8.72$  
      & $ 9.17$ & $ 7.73\pm 0.06 $ & $ 8.83 \pm 0.07$ \\
$b^-_{0+}$  & $ 1.26$ & $ 1.62$ & $ 0.82$  
      & $ 0.77$ & $ 1.56 $ & $ 0.07 \pm 0.07$ \\
$a^+_{1-}$  & $-5.52$ & $-5.38$ & $-4.90$  
      & $-5.53$ & $-5.46\pm 0.10 $ & $-5.33 \pm 0.17$ \\ 
$a^+_{1+}$  & $ 13.97$ & $13.66$ & $ 14.21$  
      & $ 13.27$ & $ 13.13\pm 0.13 $ & $ 13.6 \pm 0.1$ \\
$a^-_{1-}$  & $-1.36$ & $-1.25$ & $-0.98$  
      & $-1.13$ & $-1.19\pm 0.08 $ & $-1.00 \pm 0.10$ \\
$a^-_{1+}$  & $-8.44$ & $-8.40$ & $-8.16$  
      & $-8.13$ & $-8.22\pm 0.07 $ & $-7.47 \pm 0.13$ \\
  \hline\hline
  \end{tabular}
  \caption{Values of the S-- and P--wave threshold parameters for the various fits
           as described in the text in comparison to the respective
           data. Note that we have extracted $b_{0+}^\pm$ from the
           Matsinos phase shifts and thus no uncertainty is given. 
           Units are appropriate inverse powers of the pion mass times 10$^{-2}$.
           \label{tab:thr}}
\end{center}
\end{table}
Since we do not fit data below $q_\pi = 40, 41, 60\,$MeV (fit~1,2,3) ,
the threshold parameters are now predictions. These are
shown for the various fits in table~\ref{tab:thr} in comparison to the
empirical values of the various analyses. First, we observe that the
numbers resulting from the one--loop calculation are consistent
with the ``empirical'' ones. The latter are obtained by use of
dispersion relations for KA85 and SP98\footnote{To be precise, only the
values for $a_{0+}^\pm$ and $a_{1+}^+$ are obtained by dispersion
relations and the others by effective range fits for SP98. Some of the
results based on the effective range fits appear to be ``strange''.
We are grateful to Marcello Pavan for providing us with this information.}
 and in the EM98 case with the
help of a fairly precise tree level model~\cite{leisi}. One might, of course,
question the validity of such an approach. Notice that the isoscalar
S--wave scattering length can not be determined precisely within our
calculation (as already noted in \cite{bkmpin}) but we can give a
range, $-1.0 \leq a^+_{0+} \leq 0.6$ in units of $10^{-2}/M_{\pi^+}$.
In this band falls also the recent HBCHPT determination based on the
analysis of pion--deuteron scattering, $a^+_{0+} = (-0.30\pm 0.05) \cdot
10^{-2}/M_{\pi^+}$~\cite{silas}. For the isovector S--wave scattering 
length, we find a more precise prediction:
\beq\label{aminus}
8.3\cdot 10^{-2}/M_{\pi^+} \leq a^-_{0+} \leq 9.3\cdot
10^{-2}/M_{\pi^+}~.
\eeq
To arrive at this band, we have excluded the  
rather low value found for fit~2. This  can be traced back to 
the fact that the third order LEC contribution $\sim M_\pi^3 (\bar{d}_1 +
\bar{d}_2 + \bar{d}_3 +2\bar{d}_5)$, which is positive for fits~1,3
but is negative for fit~2 due to a negative value of $\bar{d}_5$ in that
case. This band is consistent with the one given in \cite{bkmpin2}.
In that paper, resonance saturation was used to estimate this
particular combination of LECs. While one might question the accuracy
of such an estimate, the sign should be given by the dominant $\Delta$
contribution which is positive for the off--shell parameter $Z < 0.5$.
As discussed in the review~\cite{bkmrev}, all phenomenological
indications point towards a value of $Z$ well below this limit. To
arrive at the lower end of the range given in Eq.(\ref{aminus}), we
have set the third order counterterm contribution to zero. 
Finally, we remark that due to the large uncertainty in the value of $a_{0+}^+$
one can not draw a conclusion on the value for $a_3^+ = a^+_{0+} -
a^-_{0+}$. In particular, the small value of $a_3^+  = 0.077/M_{\pi^+}$
found in ref.\cite{nadia} is consistent within the bands given above.
The resulting predictions for the S--wave scattering lengths in
comparison to the pionic atom measurements are shown in fig.~\ref{fig:apm}.
Finally, the so--called subthreshold parameters are discussed in 
app.~\ref{app:sub}.

Of course, one has to ask for the convergence. This issue is
addressed for fit~1 in fig.~7 (the results for fits~2,3 are almost identical).
In the S--waves, the second and third order contributions are small up to
$q_\pi \simeq 100\,$MeV. For higher energies, there are sizeable
cancellations between the second and the third order. 
In the P--waves, the situation is somewhat
different. While the first order describes well the partial waves
$P_{31}$ and $P_{13}$, there are sizeable second and third order
contributions of opposite sign above approximately 70~MeV. For $P_{11}$, 
the same happens even at somewhat smaller momenta.
Finally, in the case of the $P_{33}$, the third order terms
contribute very little and the tail of the $\Delta$ is encoded in the
second order contribution $\sim c_i$. For all partial waves we observe
that the counterterm contribution is much bigger than the one from
the tree graphs at that order. Note also that at third order, the loop
and the $c_i /m$ terms are much more important than the third order
counterterm amplitude $\sim d_i$. This is the reason why these third
order LECs can only be determined with a rather large uncertainty and
vary from fit to fit (although they overlap within the error
bars, cf. table~\ref{tab:LEC}). In light of these results, a fourth
order calculation is called for. This conclusion was also reached in~\cite{moj},
where the convergence of the threshold parameters was studied.

\subsection{Comparison to other calculations}

First, we come back to the LECs $c_i$. These have been obtained in~\cite{bkmlec}
by fitting threshold and subthreshold parameters as well as the $\sigma$--term. All
the quantities considered in that paper did not depend on dimension three
LECs. The value of $c_1$ we find is sizeably bigger than in~\cite{bkmlec},
leading to a $\sigma$--term of 73.0, 88.2 and 96.4~MeV for fits~1,2,3, respectively.
However, the fit to the phase shifts is not too sensitive to the
value of $c_1$. Fixing its  value as given in~\cite{bkmlec}, corresponding
to a $\sigma$--term of 47.6~MeV, the fit does only give a somewhat worse
$\chi^2$/dof for fits~1 and 3. For fit~2, the changes are more dramatic.
In all cases, it is mostly the description of the $S_{31}$ phase above 130~MeV
that worsens. Clearly, a precise determination of $c_1$ from the phase shifts 
does not seem to be possible to this order. 
Note also that an order $q^3$ calculation is certainly not sufficient
for the $\sigma$--term. This statement is underlined by the SU(3) calculation
of the baryon masses and $\sigma$--terms to order $q^4$ presented in~\cite{bm}.
The values for the LECs $c_{2,3,4}$ are in good agreement with the values
found in~\cite{bkmlec}, only $c_2$ comes out slightly larger here. If one,
however, fixes the values of the LECs to the previously determined values,
the fits get clearly worse. We believe that the method employed here,
namely to fit to various sets of phase shifts, gives a better handle on the
theoretical uncertainties than the method used in~\cite{bkmlec}. A more consistent
picture of the LEC $c_1$ will certainly emerge in a calculation at order $q^4$,
which goes beyond the scope of this paper. Notice also that the combination
$c^+ =c_2+c_3-2c_1$, which enters directly the formula for the isoscalar
S--wave scattering length, is very small in all cases. We find $c^+ =
-0.20,+0.13,+0.08\,$GeV$^{-1}$  for fits~1,2,3, respectively and of the same
size than in~\cite{bkmlec,silas}, where $c^+ = -0.09\pm 0.37\,$GeV$^{-1}$.

The results of ref.\cite{moj} are comparable for the LECs $c_i$ but the
resulting phase shifts deviate earlier from the data as shown in ref.~\cite{dapa}.
This underlines our point that using only threshold parameters to fix the
LECs is a less precise method. A better description of the phase shifts
also at higher energies is, of course, achieved when one introduces the
$\Delta$--resonance explicitely in the effective Lagrangian. The most elaborate
calculation in this respect is the one of Ellis and Tang~\cite{elta1,elta2}.
They employ a different counting scheme by expanding the relativistic pion--nucleon
loop graphs. While this procedure might be legitimate at order $q^3$, it is not
clear to us how one can systematically extend this scheme to higher orders since
at some point it will face the same difficulties encountered by 
Gasser et al.~\cite{gss} in the fully relativistic calculation. Furthermore, the
inclusion of the $\Delta$--isobar is not done along the lines of a consistent 
power counting as outlined recently by Hemmert et al.~\cite{hhk}. The
full complexity of consistently including the $\Delta$ can be seen
from the renormalization of $g_A$ as spelled out in~\cite{bfhm}.
While the
approach of~\cite{elta2} is clearly superior for describing the phases up to
higher energies, the amplitude we have constructed can certainly be controlled
better in the extrapolation to and below the threshold. Furthermore, it can
immediately be extended to include virtual photons~\cite{ms}. The issue of
isospin violation as evaluated in the consistent CHPT machinery, i.e. in an
effective field theory including pions, nucleons and virtual photons, 
will be addressed in a forthcoming publication.

\section*{Acknowledgements}

We are grateful to Gerhard H\"ohler, Evangelos Matsinos and Marcello
Pavan  for useful communications.

\bigskip
\appendix
\section{Field transformations and the minimal Lagrangian}
\def\theequation{\Alph{section}.\arabic{equation}}
\setcounter{equation}{0}
\label{app:trafo}

In this appendix, we will show how one can use field transformations
to construct the minimal Lagrangian discussed in section~2. We consider
relativistic spin-1/2 fields since we always construct first the minimal
Lagrangian before performing the heavy fermion limit. Stated differently, our starting
point is always the relativistic pion--nucleon theory which is mapped onto
the heavy fermion limit by means of the path integral as spelled out in
\cite{bkkm}. 

First, we will be concerned with the reduction of terms which are proportional
to the mesonic equations of motion (eom) derived from the lowest order (dimension two)
effective meson Lagrangian (note, however, that one can use this method
also at higher orders, i.e. one does not have to start with the mesonic eom
derived from ${\cal L}^{(2)}_{\pi\pi}$). The mesonic eom used here reads
\beq
2 \, [iD_\mu, u^\mu ] = \chi_- - \frac{1}{2} \langle \chi_- \rangle \,\,.
\eeq
We will show now how one can systematically  eliminate such terms. As an example,
consider the possible scalar--isoscalar combination contributing to the effective
pion--nucleon Lagrangian,
\beq\label{term}
\bar{\Psi} \, \gamma_5 \, [iD_\mu, u^\mu ]\, \Psi\,\,\  .
\eeq
Such type of terms do not appear in the minimal form of the Lagrangian
given in Eq.(\ref{LpiNrel}).
To eliminate such a contribution, we redefine the $U$--field in the following way
(note that $U'$ denotes the field before the transformation subject to the
conditions $(U')^\dagger U' = 1$ and det~$U'=1$):
\beq
U = \exp \biggl\lbrace -\frac{i}{2} \bar{\Psi} i \gamma_5 \tau^i \Psi 
\, u \tau^i u^\dagger \biggr\rbrace \, U' \,\, .
\eeq 
It is easy to check that the argument of the exponential is hermitean,
i.e. we are dealing with a unitary transformation, $U^\dagger U = UU^\dagger =1$.
Furthermore, the property det~$U=1$ has to be preserved. This follows
from
\beqa
{\rm det}~U &=& {\rm det}~\biggl( \exp \biggl\lbrace -\frac{i}{2} \bar{\Psi} 
i \gamma_5 \tau^i \Psi  \, u \tau^i u^\dagger \biggr\rbrace \biggr) \,
{\rm det}~U' \nonumber \\
&=& \exp \biggl\lbrace -\frac{i}{2} \bar{\Psi} i \gamma_5 \tau^i \Psi 
\, \langle u \tau^i u^\dagger \rangle \biggr\rbrace \nonumber \\
&=& {1} \,\,\, ,
\eeqa
using $ \langle u \tau^i u^\dagger \rangle = 0$ and det~$U'=1$. Consequently, the
lowest order meson Lagrangian takes the form
\beqa
{\cal L}_{\pi\pi}^{(2)} &=& \frac{F^2}{4} \langle D_\mu U D^\mu U^\dagger 
+ \chi \, U^\dagger + U \, \chi^\dagger \rangle  \\
 &=& \frac{F^2}{4} \langle D_\mu U' D^\mu (U')^\dagger 
+ \chi \, (U')^\dagger + U' \, \chi^\dagger \rangle
+ {\cal L}^{(1)}_{\bar{\Psi}\Psi, {\rm induced}}
+ {\cal L}^{(1)}_{\bar{\Psi}\Psi\bar{\Psi}\Psi, {\rm induced}} + \ldots~.
\nonumber
\eeqa 
Note that such type of field redefinitions induce by construction a whole string with 
two, four, $\ldots$ nucleon fields chirally coupled to pions and external sources. 
Here we are only interested in processes with
exactly one nucleon line running through the pertinent Feynman diagrams. In 
this case, the induced term reads
\beqa\label{induced}  
{\cal L}^{(1)}_{\bar{\Psi}\Psi, {\rm induced}} &=& i \frac{1}{2} \bar{\Psi}
i \gamma_5  \tau^i \Psi \langle \tau^i \chi_- \rangle+ i \frac{1}{2} \bar{\Psi}
i \gamma_5  \tau^i \Psi \langle \tau^i [iD_\mu , u^\mu] \Psi \nonumber \\
&=& -\bar{\Psi}  \gamma_5 \chi_- \Psi + \frac{1}{2} \bar{\Psi}  \gamma_5 
\langle \chi_- \rangle \Psi - \bar{\Psi} \gamma_5  [iD_\mu , u^\mu] \Psi~.
\eeqa
The last term in Eq.({\ref{induced}) obviously cancels the one given in
Eq.(\ref{term}). The other terms in Eq.({\ref{induced})
can be eliminated by applying similar field transformations on the nucleon
fields as discussed below. The multi--fermion terms not discussed here play, 
of course, a role in the construction of chiral invariant few--nucleon forces.

Consider now the nucleon eom terms. Be $A$ an arbitrary local operator
of chiral dimension two (or higher). We redefine the nucleon spinor via
\beq
\Psi = (1+\gamma_0 A^\dagger \gamma_0)\,\Psi ' \,\, , \quad
\bar{\Psi} =\bar{\Psi} ' (1+A) \,\,\, .
\eeq
This induces a string of additional terms in the leading order
pion--nucleon Lagrangian,
\beqa
{\cal L}_{\pi N}^{(1)} &=& {\cal L}_{\pi N}^{(1\, ')}  
+ {\cal L}_{\pi N, {\rm induced}}^{(1)} \\
{\cal L}_{\pi N, {\rm induced}}^{(1)} &=& \bar{\Psi} \biggl( -i 
\stackrel{\leftarrow}{\barr D} \gamma_0 A^\dagger \gamma_0 + 
A i \stackrel{\rightarrow}{\barr D} \biggr) \Psi
-m\bar{\Psi} (A + \gamma_0 A^\dagger \gamma_0) \Psi \nonumber \\
&+&\frac{g_A}{2} \bar{\Psi} (A \barr u \gamma_5 + \barr u \gamma_5
\gamma_0 A^\dagger \gamma_0 ) \Psi.
\eeqa
As an example, consider the third order term
\beq\label{lpinex} 
{\cal L}_{\pi N}^{(3)} = \bar\Psi \gamma_5 \chi_- \Psi + \ldots \,\,\, ,
\eeq
which is exactly the structure that appeared before. With $A= \gamma_5 
\chi_- / (2m)$, we have
\beq
\Psi = \biggl(1+ \frac{1}{2m} \gamma_5 \chi_-\biggr)\,\Psi ' 
\,\, , \quad
\bar{\Psi} =\bar{\Psi} ' \biggl( 1+  \frac{1}{2m} \gamma_5 \chi_-
\biggr) \,\,\, ,
\eeq
so that
\beqa
{\cal L}_{\pi N, {\rm induced}}^{(3)} &=& \frac{1}{2m} \bar{\Psi}'
\biggl( -i \stackrel{\leftarrow}{\barr D} \gamma_5 \chi_- + 
\gamma_5 \chi_- i \stackrel{\rightarrow}{\barr D} \biggr) \Psi ' 
-\bar{\Psi}' \gamma_5 \chi_- \Psi - \frac{g_A}{4m} \bar{\Psi}' \gamma_\mu
[ \chi_- , u^\mu ] \Psi ' \nonumber \\
&=&
- \bar{\Psi}' \gamma_5 \chi_- \Psi + \frac{1}{2m} \bar{\Psi}' \gamma_\mu
\gamma_5 [D_\mu , \chi_- ] \Psi '  - \frac{g_A}{4m} \bar{\Psi}' \gamma_\mu
[ \chi_- , u^\mu ] \Psi ' \,\,\, .
\eeqa
Therefore, the term in Eq.(\ref{lpinex}) is cancelled by the first term
of the induced Lagrangian.
Notice that this is equivalent to the  relation~Eq.(\ref{PI1}) but derived
simply by field redefinitions. In a similar fashion, one can construct
other operators $A$ which then lead to all other relations given in
Eqs.(\ref{R1}-\ref{PI2}).

\section{Reduction of the Clifford--algebra in HBCHPT}
\def\theequation{\Alph{section}.\arabic{equation}}
\setcounter{equation}{0}
\label{app:cliff}

In this appendix, we collect the various expressions of the $\gamma$
matrices expressed in terms of the nucleon four--velocity and the
covariant spin--operator. Consider first $\bar{H} H $ matrix elements:
\begin{eqnarray}
\bar{H} H & = & \bar{H} H~, \\
\bar{H} \gamma_\mu H & = & v_\mu \bar{H} H~, \\
\bar{H} \gamma_5 H & = & 0~, \\
\bar{H} \gamma_\mu \gamma_5 H & = & 2\,\bar{H} S_\mu H~, \\
\bar{H} \sigma_{\mu\nu} H & = & 2\,\epsilon_{\mu\nu\alpha\beta}
v^\alpha \bar{H} S^\beta H~, \\
\bar{H} \gamma_5 \sigma_{\mu\nu} H & = & 2i\,\bar{H} 
(v_\mu S_\nu - v_\nu S_\mu) H~. 
\end{eqnarray}
Similarly, $\bar{h} H $ matrix elements reduce to
\begin{eqnarray}
\bar{h} H & = & 0~, \\
\bar{h} \gamma_\mu H & = & \bar{h} (\gamma_\mu - v_\mu) H~, \\
\bar{h} \gamma_5 H & = & \bar{h} \gamma_5 H~, \\
\bar{h} \gamma_\mu \gamma_5 H & = & -v_\mu \bar{h} \gamma_5 H~, \\
\bar{h} \sigma_{\mu\nu} H & = & -i\,\bar{h} (v_\mu \gamma_\nu 
      - v_\nu \gamma_\mu) H~, \\
\bar{h} \gamma_5 \sigma_{\mu\nu} H & = & 
-\epsilon_{\mu\nu\alpha\beta} v^\alpha \bar{h} \gamma^\beta H~.
\end{eqnarray}
The ones for $\bar{H} h$ are identical, only the last three expressions
have a plus sign on the right hand side.
Finally, we list some of the operators which appear when
performing  the $1/m$ expansion of the generating functional:
\begin{eqnarray}
\bar{H} (\gamma_\mu - v_\mu) (\gamma_\nu -v_\nu) H 
& = & \bar{H} (g_{\mu\nu} - v_\mu v_\nu -2 [S_\mu,S_\nu]) H~, \\
\bar{H} (\gamma_\mu - v_\mu) \gamma_5 H & = & 2\,\bar{H} S_\mu H~, \\
\bar{H} \gamma_5 (\gamma_\mu - v_\mu) H & = & -2\,\bar{H} S_\mu H~, \\
\bar{H} (\gamma_\mu - v_\mu) S_\nu \gamma_5 H
& = & -1/2\,\bar{H} (g_{\mu\nu} - v_\mu v_\nu - 2[S_\mu,S_\nu]) H~, \\
\bar{H} \gamma_5 S_\nu (\gamma_\mu - v_\mu) H 
& = & 1/2\,\bar{H} (g_ {\mu\nu} - v_\mu v_\nu + 2[S_\mu,S_\nu]) H~, \\
\bar{H} (\gamma_\mu - v_\mu) S_\nu (\gamma_\lambda - v_\lambda) H 
& = & \bar{H} (-i/2\,\epsilon_{\mu\nu\lambda\alpha} v^\alpha 
+ (g_{\mu\nu} - v_\mu v_\nu) S_\lambda \nonumber\\
& & - (g_{\mu\lambda} - v_\mu v_\lambda) S_\nu + (g_{\nu\lambda} 
- v_\nu v_\lambda) S_\mu) H~.
\end{eqnarray}


\section{Feynman rules from the third order Lagrangian}
\def\theequation{\Alph{section}.\arabic{equation}}
\setcounter{equation}{0}
\label{app:rules}

In this appendix, we give the Feynman rules needed to work out
$\pi N$ scattering based on the Lagrangian, Eq.(\ref{lpin3f}). 
We use the following conventions: The momenta $p_1$ and $q_1$ are always
{\it in}coming, whereas $p_2$, $q_2$ and $q$ (the latter is the
pion momentum in case of the $\pi N$ vertex) are always {\it
  out}going. Also, '$a,b,c$' are pion isospin indices.
Consider first the contributions from the terms
with {\it fixed} coefficients (this includes the terms
which are $1/m$--corrections to  dimension two operators with
LECs $c_i$):

\medskip

\noindent Nucleon kinetic energy
\beq
i \frac{(v \cdot p_1)^3 - (v \cdot p_1) \, p_1^2}{4 m^2}~;
\eeq
one pion 
\beqa
-\frac{g_A}{4m^2 F} &&\!\!\!\!\! \biggl\{\frac{1}{2}  S\cdot q q ^2 
+ S\cdot q   v \cdot p_1 v \cdot p_2 - v \cdot (p_1+p_2)
S \cdot (p_1+p_2) v \cdot q  \nonumber \\
&& - \frac{1}{2} S\cdot q   (p_1^2 + p_2^2) + [ S\cdot p_2 q \cdot p_1
+ S\cdot p_1 q \cdot p_2 ]  \biggr\}\, \tau^a~;
\eeqa
two pions
\beqa
  i\, \frac{\delta^{ab}}{F^2} && \!\!\!\!\! \biggl\{
\frac{g_A^2}{16m^2}[3v\cdot q_1 v\cdot q_2 v \cdot(p_1+p_2)
- v\cdot q_1 q_2\cdot(p_1+p_2) -  v\cdot q_2 q_1\cdot(p_1+p_2)]
\nonumber \\ &&
+ \frac{c_2}{m}[ v\cdot q_1 q_2\cdot(p_1+p_2)+ v\cdot q_2
q_1\cdot(p_1+p_2)] \nonumber \\ &&
+ \frac{g_A^2}{8 m^2}[S\cdot q_1 , S\cdot q_2 ] 
v \cdot (q_1+q_2) \biggr\} \nonumber \\
-\frac{\epsilon^{bac}\, \tau^c}{8m^2F^2} && \!\!\!\!\! \biggl\{
\frac{1}{2} v\cdot(q_1+q_2) \bigl[
(v\cdot p_1)^2 + (v\cdot p_2)^2 + v\cdot p_1 v\cdot p_2 \bigr]
\nonumber\\ && 
- \frac{1}{4} \bigl[ (p_1^2 + p_2^2) v\cdot(q_1+q_2)
+ v \cdot(p_1+p_2) (p_1+p_2)\cdot (q_1+q_2) \bigr]
\nonumber \\ &&
+ [S\cdot q_1 , S\cdot q_2 ] \, v\cdot (p_1+p_2) \nonumber \\ &&
+ (1+g_A^2+8mc_4)\, [ v \cdot q_2 S\cdot q_1 - v \cdot q_1 S \cdot
q_2, S \cdot (p_1+p_2)] \nonumber \\ &&
-\frac{g_A^2}{2}[v \cdot q_2 q_1 (p_1-p_2) + v\cdot q_1 q_2
(p_2 -p_1) -v\cdot q_1 v\cdot q_2 v \cdot (q_1+q_2) ]\nonumber \\ &&
+  \frac{1}{2} (1+8mc_4) \, [  v\cdot q_1 q_2^2 + v \cdot q_2 q_1^2
- q_1 q_2 v \cdot (q_1+q_2)] \biggr\} \,\, .
\eeqa
Similarly, for the terms with LECs from ${\cal L}_{\pi N}^{(3)}$, we get:

\medskip

\noindent Nucleon kinetic energy
\beq
-i \tilde{d}_{24} (\lambda ) (v \cdot p_1)^3 
+ 8i M^2 \tilde{d}_{28} (\lambda ) (v \cdot p_1)~;
\eeq
one pion 
\beqa 
&& \frac{2M^2}{F} [2 {d}_{16} (\lambda ) - {d}_{18}]
S \cdot q \tau^a \nonumber \\
&& + \frac{1}{F} [\tilde{d}_{25}(\lambda ) v\cdot p_1 v\cdot p_2 
+ \tilde{d}_{29}(\lambda ) ( v\cdot q)^2] S\cdot q \tau^a~; 
\eeqa
two pions
\beqa
  i\, \frac{2\delta^{ab}}{F^2} && \!\!\!\!\! \biggl\{\bigl[
-d_{14}(\lambda ) + d_{15} \bigr]
[S\cdot q_1 , S\cdot q_2 ] v \cdot (q_1+q_2) \nonumber \\ &&
+ 2 [ \tilde{d}_{26}  (\lambda ) q_1\cdot q_2  
+ \tilde{d}_{27} (\lambda ) v\cdot q_1 v\cdot q_2 
- \tilde{d}_{28}  (\lambda ) M^2 ] v\cdot (p_1+ p_2) \biggr\}
\nonumber \\
-\frac{2\epsilon^{bac}\, \tau^c}{F^2} && \!\!\!\!\! \biggl\{
\bigl( d_1 (\lambda) + d_2 (\lambda ) 
\bigr)  q_1 \cdot q_2 +  d_3 (\lambda) \,  v\cdot q_1 v \cdot q_2
\nonumber \\ &&
+ \biggl[ 2M^2  d_5 (\lambda) - \frac{1}{8}\tilde{d}_{24} (\lambda ) 
[(v\cdot p_1)^2 + (v\cdot p_2)^2 + v\cdot p_1 v\cdot p_2] \nonumber\\&& 
+ M^2 \tilde{d}_{28} (\lambda ) \biggr] v\cdot (q_1+q_2)+ 
2 \tilde{d}_{30} (\lambda )
[S\cdot q_1 , S\cdot q_2 ] \, v\cdot (p_1+p_2) \biggr\}~.
\eeqa

\section{Kinematics of pion--nucleon scattering}
\def\theequation{\Alph{section}.\arabic{equation}}
\setcounter{equation}{0}
\label{app:kinematics}

We consider the $\pi N$-scattering amplitude without assuming isospin
symmetry (which is necessary if one wants to address isospin 
violation)\footnote{The reduction of these formulae to the
  isospin--conserving case is obvious.}
\beq
\pi (q_1) + N(p_1) \to \pi (q_2) + N(p_2)~.
\eeq
Here $\pi (q_i)$ denotes a pion
state in the physical basis $(\pi^0, \pi^+ , \pi^- )$ with
four--momentum $q_i$ and $N$ a proton ($p$)
or a neutron ($n$) with four--momentum $p_i$. 
The masses of the in(out)--going nucleon and pion are
denoted by $m_{1(2)}$ and  $M_{1(2)}$, respectively.
Consider the center-of-mass system (cms)  with ${\vec p}_i
= -{\vec q}_i$, $i=1,2$. For the initial nucleon in the heavy fermion approach,
we set $p_1 = m_1 \, v_1 + k_1$ with $k_1 \cdot p_1 \ll m_1$.
The pion and the nucleon energy in the in--state are (we use $v_1 =
(1,0,0,0)$) 
\beq
\omega_1 = v_1 \cdot q_1 = \sqrt{M_1^2 + {\vec
q\,}_1^2} \,\, , \quad E_1 = \sqrt{m_1^2 + \omega_1^2 - M_1^2}~,
\eeq
in order. The energy of the out--going pion and nucleon reads
\beq
\omega_2  = {(E_1 + \omega_1 )^2 + M_2^2 - m_2^2 \over
2 (E_1 + \omega_1 )} \,\, , \quad E_2 = E_1 + \omega_1 - \omega_2
\,\,\, ,
\eeq
respectively.  Furthermore, the invariant momentum squared is
\beqa
t = (q_1 -q_2)^2 &=& M_1^2 + M_2^2 - 2\omega_1 \omega_2 + 2{\vec q}_1 \cdot
{\vec q}_2 \nonumber \\
&=& M_1^2 + M_2^2 - 2\omega_1 \omega_2 + 2|{\vec q}_1| 
|{\vec q}_2| \cos(\theta)~,
\eeqa
and $\cos(\theta) \equiv z$ in the following. 
For the equal mass case ($M_1=M_2=M_\pi, \, m_1=m_2=m_N$), 
it is convenient to use the pion kinetic energy in the cms
given by 
\beq
T_\pi = \omega_{\rm cms} - M_\pi \quad ,
\eeq
and the pion three--momentum in the lab system, called
$\vec{q}_\pi$, with norm $q_\pi$,
\beq
q_\pi = \sqrt{ \frac{1}{4m^2} \bigl(s-M_\pi^2-m_N^2\bigr)^2 -
M_\pi^2}~.
\eeq 
For example, $T_\pi = 50\,$MeV is equivalent to $q_\pi  = 155.4\,$MeV. 
For later use, we also need the following products involving the
nucleon spinors (again written for the general case)
\beqa
\bar u (p_2) \, u(p_1) &=& {\cal N}_1  {\cal N}_2 \Bigg[ 1 
- \frac{{\vec q}_1 \cdot {\vec q}_2}{(E_1+m_1)(E_2+m_2)} + 
\frac{i\sigma\cdot({\vec q}_1 \times {\vec
  q}_2)}{(E_1+m_1)(E_2+m_2)}\Bigg]~,\\
\bar u (p_2) \barr{q}_1 u(p_1) &=& {\cal N}_1 {\cal N}_2
\bigg[ \omega_1 + \frac{ {\vec q\,}_1^2}{E_1+m_1} 
+ \frac{{\vec q}_1 \cdot {\vec q}_2}{E_2+m_2}\bigg(1 
+ \frac{\omega_1}{E_1+m_1}\bigg) \nonumber \\
&& \qquad\qquad\qquad\qquad - \frac{i\sigma\cdot({\vec q}_1 \times {\vec
  q}_2)}{E_2+m_2}\bigg(1 + \frac{\omega_1}{E_1+m_1}\bigg)\Bigg]~,
\eeqa
where the ${\cal N}_i$ are the conventional normalization factors,
\beq\label{spinor}
{\cal N}_i = \sqrt{\dfrac{E_i+m_i}{2m_i}} \,\, , \quad i =1,2~.
\eeq

\section{Subthreshold parameters}
\def\theequation{\Alph{section}.\arabic{equation}}
\setcounter{equation}{0}
\label{app:sub}

In this appendix, we consider the expansion of the standard invariant
pion--nucleon amplitudes with the pseudovector Born terms subtracted
(as indicated by the ``bar'')
\beq
\bar{X} = \sum_{m,n} x_{mn} \,\nu^{2m+k} \, t^n \,\, , \quad
X = \{A^\pm, B^\pm \} \,\, ,
\eeq
with 
\beq
\nu = \frac{t}{4m} + \frac{1}{m} \biggl( \omega^2 - M^2 + \omega
\sqrt{m^2 + \omega^2 -M^2} \biggr) \,\, ,
\eeq
and $k = 1 \,(0)$ if the function considered is odd (even) in $\nu$.
Instead of the amplitudes $B^\pm$, we choose to work with $D^\pm =
A^\pm + \nu\,B^\pm$. In particualr, the $B^+$ amplitude is dominated
by the $\Delta (1232)$ and can only be expected to described well in a
fourth order calculation. The invariant amplitudes $A^\pm$ and $B^\pm$ can be
reconstructed from the $g^\pm, h^\pm$ amplitudes used throughout the
main body of the paper via
\beq
A^\pm = C_1 \, g^\pm + C_2 \, h^\pm \,\, , \quad
B^\pm = C_3 \, g^\pm + C_4 \, h^\pm \,\, , 
\eeq
with
\beqa
C_1 &=& \frac{\sqrt{s}+m}{2\sqrt{s}} \,\, , \,\, 
C_2  = -\frac{1}{2\sqrt{s}} \biggl( (\omega^2 - M^2 )
\frac{2E\sqrt{s}-2m^2}{E-m} + (\sqrt{s}+m) \frac{t}{2} \biggr)~,
\nonumber\\
C_3 &=& \frac{1}{2\sqrt{s}} \,\, , \,\, 
C_4  =  \frac{1}{2\sqrt{s}} \biggl( (\omega^2 - M^2 )
\frac{2m}{E-m} -  \frac{t}{2} \biggr) \,\, .
\eeqa
It is now straightforward to work out the subthreshold expansion
around $\nu = t = 0$.
Here, we collect the results for the coefficients of the subthreshold
expansion  which involve only lower powers in $t$ and/or $\nu$ and are
accessible in a $q^3$ calculation (see also app.~A of~\cite{bkmlec}
for some additional results). 
\begin{table}[hbt]
\begin{center}
\begin{tabular}{|c|c|c|c|c|}
    \hline\hline
    Parameter  & Fit~1   &  Fit~2 & Fit~3 & KH-analysis \\
    \hline
$a^+_{00} [M_\pi^{-1}]$ & $-1.57$ & $-1.27$ & $-1.35$ & 
                          $-1.46\pm 0.10$  \\
$a^+_{10} [M_\pi^{-3}]$ & $ 9.10$ & $ 9.06$ & $ 9.17$ & 
                          $ 4.66$  \\
$a^+_{01} [M_\pi^{-3}]$ & $ 1.37$ & $ 1.34$ & $ 1.45$ & 
                          $ 1.14\pm 0.02$  \\
$a^+_{11} [M_\pi^{-5}]$ & $ 0.07$ & $ 0.07$ & $ 0.07$ & 
                          $-0.01$  \\
$d^+_{10} [M_\pi^{-3}]$ & $ 1.23$ & $ 1.14$ & $ 1.19$ & 
                          $ 1.12\pm 0.02$  \\
$d^+_{01} [M_\pi^{-3}]$ & $ 1.37$ & $ 1.34$ & $ 1.45$ & 
                          $ 1.14\pm 0.02$  \\
$a^-_{00} [M_\pi^{-2}]$ & $-8.47$ & $-8.88$ & $-8.73$ & 
                          $-8.83\pm 0.10$  \\
$a^-_{10} [M_\pi^{-4}]$ & $-1.45$ & $-1.36$ & $-1.43$ & 
                          $-1.25\pm 0.05$  \\
$a^-_{01} [M_\pi^{-4}]$ & $-0.46$ & $-0.48$ & $-0.43$ & 
                          $-0.37\pm 0.02$  \\
$a^-_{11} [M_\pi^{-6}]$ & $ 0.0 $ & $ 0.0 $ & $ 0.0 $ & 
                          $ 0.01\pm 0.01$  \\
$d^-_{00} [M_\pi^{-2}]$ & $ 1.87$ & $ 1.59$ & $ 1.80$ & 
                          $ 1.53\pm 0.02$  \\
$d^-_{10} [M_\pi^{-4}]$ & $-0.65$ & $-0.56$ & $-0.63$ & 
                          $-0.17\pm 0.01$  \\
$d^-_{01} [M_\pi^{-4}]$ & $-0.32$ & $-0.34$ & $-0.29$ & 
                          $-0.13\pm 0.01$  \\
   \hline\hline
  \end{tabular}
\caption{Subthreshold parameters for the $A^\pm$ and $D^\pm$
    amplitudes for the three fits in comparison to the 
    Karlsruhe-Helsinki analysis.}\label{tab:sub}
\end{center}
\end{table}
We find (we only give the formulas for the subthreshold parameters
not already given in~\cite{bkmlec}):
\beqa
a_{10}^+ &=& \frac{1}{8F_\pi^4 \pi^2} \biggl[ 16F_\pi^2 \pi^2 c_2 -
32 m F_\pi^2
\pi^2 (\bar{d}_{14} - \bar{d}_{15}) + mg_A^4 \biggr] - \frac{M_\pi}{8\pi
  F_\pi^4} \biggl[ \frac{5g_A^2}{4} + 1 \biggr]~,  \\
a_{00}^- &=& -\frac{2mc_4}{F_\pi^2} + \frac{M_\pi m g_A^2
  (1+g_A^2)}{8\pi F_\pi^4} \nonumber\\
&+& \frac{M_\pi^2}{48m^2F_\pi^4\pi^2}\biggl[ g_A^4 m^2 + 6F_\pi^2
\pi^2 (1 - 4 mc_4 + 32m^2 (\bar{d}_1+\bar{d}_2) + 64 m^2 \bar{d}_5 )
\biggr]~, \\
a_{10}^- &=& -\frac{m g_A^4}{32\pi F_\pi^4 M_\pi} 
- \frac{1}{240 m^2 F_\pi^4 \pi^2}\biggl[ (15+7g_A^4) m^2 + 30F_\pi^2
\pi^2 (1 - 4 mc_4 + 32m^2 \bar{d}_3  )\biggr]~,\nonumber \\ && \\
a_{01}^- &=& -\frac{m g_A^2}{96\pi F_\pi^4 M_\pi} 
- \frac{1}{192 m F_\pi^4 \pi^2}\biggl[ (1+7g_A^2+2g_A^4) m + 48F_\pi^2
\pi^2 (-c_4 + 8m (\bar{d}_1+\bar{d}_2)  )\biggr]~, \nonumber\\ &&\\
d_{00}^- &=& \frac{1}{2F_\pi^2} + \frac{M_\pi^2}{48\pi^2 F_\pi^4} 
\biggl[ g_A^4  + 192 F_\pi^2 \pi^2 (\bar{d}_1+\bar{d}_2 +2 \bar{d}_5 )
\biggr]~, \\
d_{10}^- &=& -\frac{1}{240\pi^2 F_\pi^4} 
\biggl[ 15+7g_A^4 - 960 F_\pi^2 \pi^2 \bar{d}_3\biggr]~, \\
d_{01}^- &=& -\frac{1}{192\pi^2 F_\pi^4} 
\biggl[ 1+7g_A^2+2g_A^4 + 384 F_\pi^2 \pi^2 (\bar{d}_1
+\bar{d}_2)\biggr]~.
\eeqa
Notice that $d_{0n}^+ = a_{0n}^+$ for any $n$. The resulting values
are given in table~\ref{tab:sub}. The ``empirical'' values based 
on a dispersion--theoretical continuation into the unphysical region 
are taken from \cite{hoeh}. For the VPI SP98 analysis, we have in
addition~\cite{SAID}
\beq\label{subp}
d_{00}^+ = -1.30 \, M_\pi^{-1} \,\, , \quad 
d_{01}^+ =  1.27 \, M_\pi^{-3} \,\, .
\eeq 
Obviously, only in some cases the one-loop result is in good agreement with
the empirical values as deduced from the Karlsruhe--Helsinki (KH) phase shift 
analysis. Note, however, that recent low energy $\pi N$-scattering data 
from PSI \cite{joram} (see also ref.\cite{me97})
show some disagreement with the KH80 solution of $\pi N$ dispersion 
analysis. It therefore seems necessary to redo the $\pi N$-dispersion analysis
with the inclusion of these new data. This was done partly in the VPI
SP98 analysis leading e.g. to the values of the two subthreshold
parameters quoted in Eq.(\ref{subp}). However, a full--scale
dispersive reanalysis for all (sub)threshold parameters is not yet available.

\bigskip

\newpage

\noindent {\Large {\bf Figures}}

$\,$

\vskip 1.5cm

\begin{figure}[hbt]
\centerline{
\epsfysize=1.2in
\epsffile{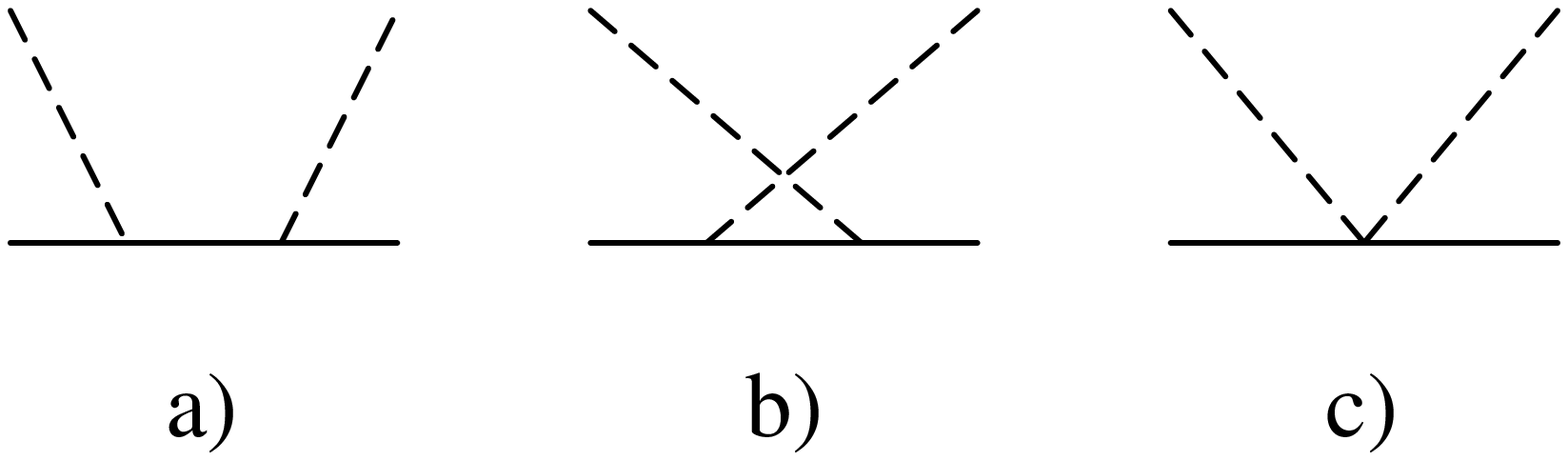}
}
\vskip 0.4cm
\caption{Tree graphs in the relativistic approach as discussed in the
  text.}

\end{figure}

$\,$

\vskip 1cm

\begin{figure}[bht]
\centerline{
\epsfysize=3.7in
\epsffile{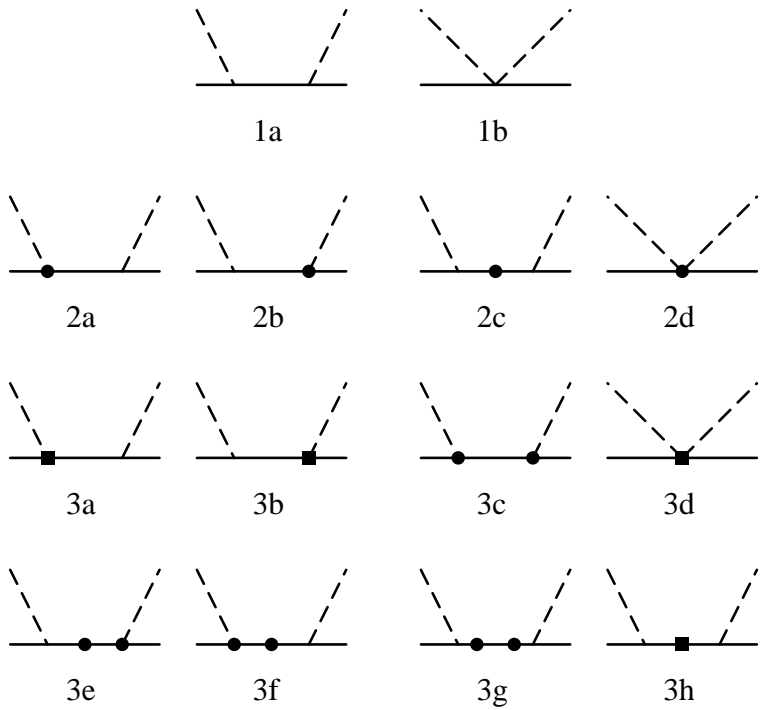}
}
\vskip 0.5cm
\caption{Tree graphs in the heavy baryon approach. The heavy dots
  refer to insertions from ${\cal L}^{(2)}_{\pi N}$ and the filled
  squares to insertions from ${\cal L}^{(3)}_{\pi N}$.}

\end{figure}

\newpage

\vskip 1cm

\begin{figure}[bht]
\centerline{
\epsfysize=7in
\epsffile{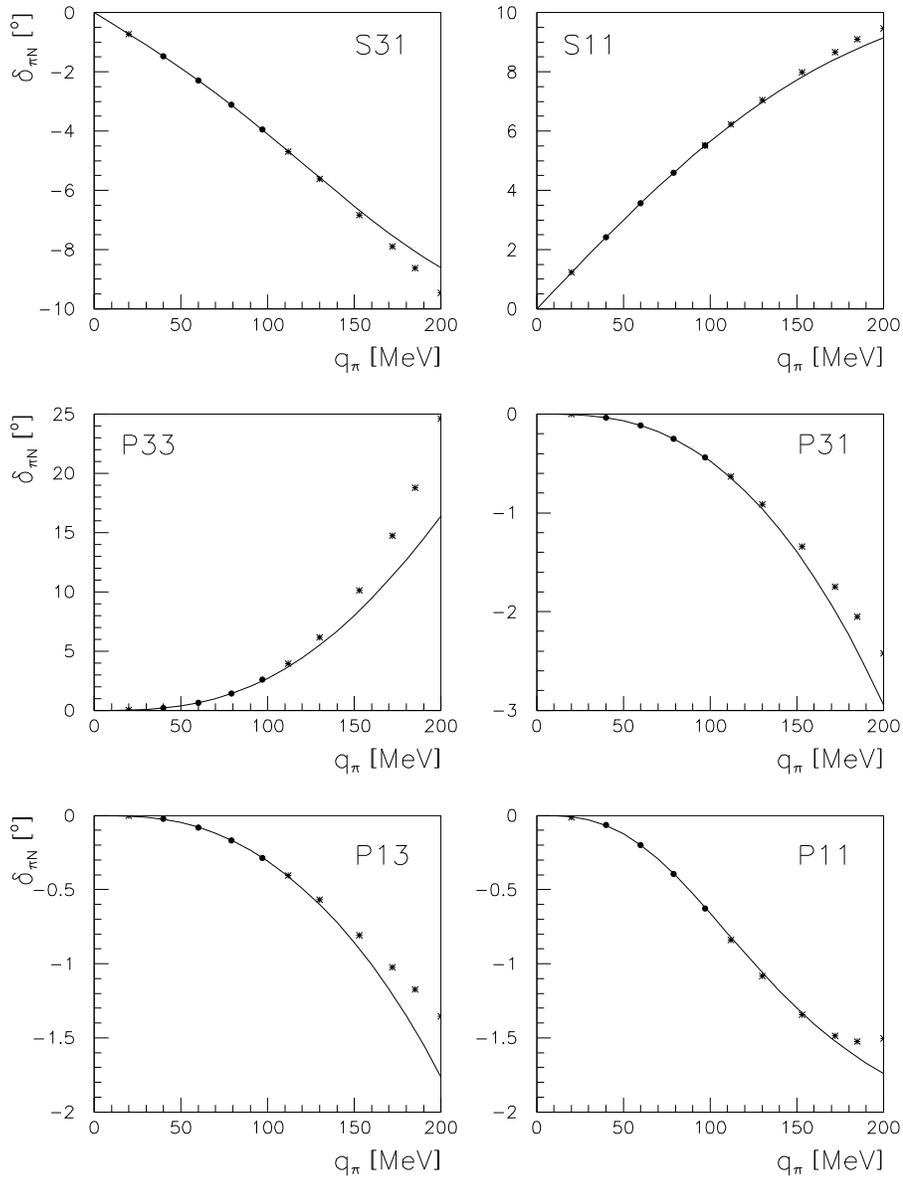}
}
\vskip 0.5cm
\caption{
Fits and predictions for the KA85 phase shifts as a
function of the pion laboratory  momentum $q_\pi$. Fitted in
each partial wave are the data between 40 and 97~MeV (filled
circles). For higher
and lower energies, the phases are predicted.
}

\end{figure}

\newpage

\vskip 1cm

\begin{figure}[bht]
\centerline{
\epsfysize=7in
\epsffile{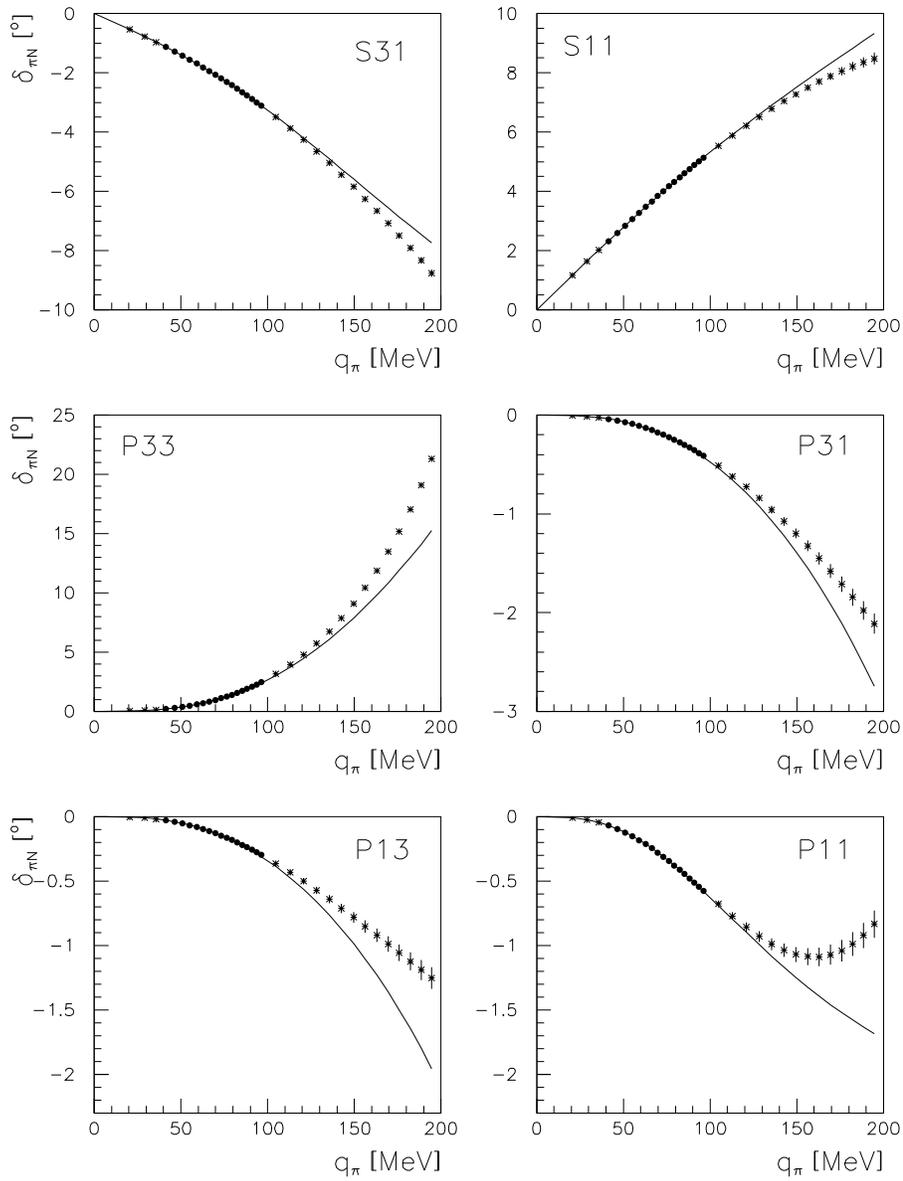}
}
\vskip 0.5cm
\caption{
Fits and predictions for the EM98 phase shifts  as a
function of $q_\pi$. Fitted in
each partial wave are the data between 41 and 97~MeV (filled circles). For higher
and lower energies, the phases are predicted as shown by the solid lines.
}

\end{figure}

\newpage

\vskip 1cm

\begin{figure}[bht]
\centerline{
\epsfysize=7in
\epsffile{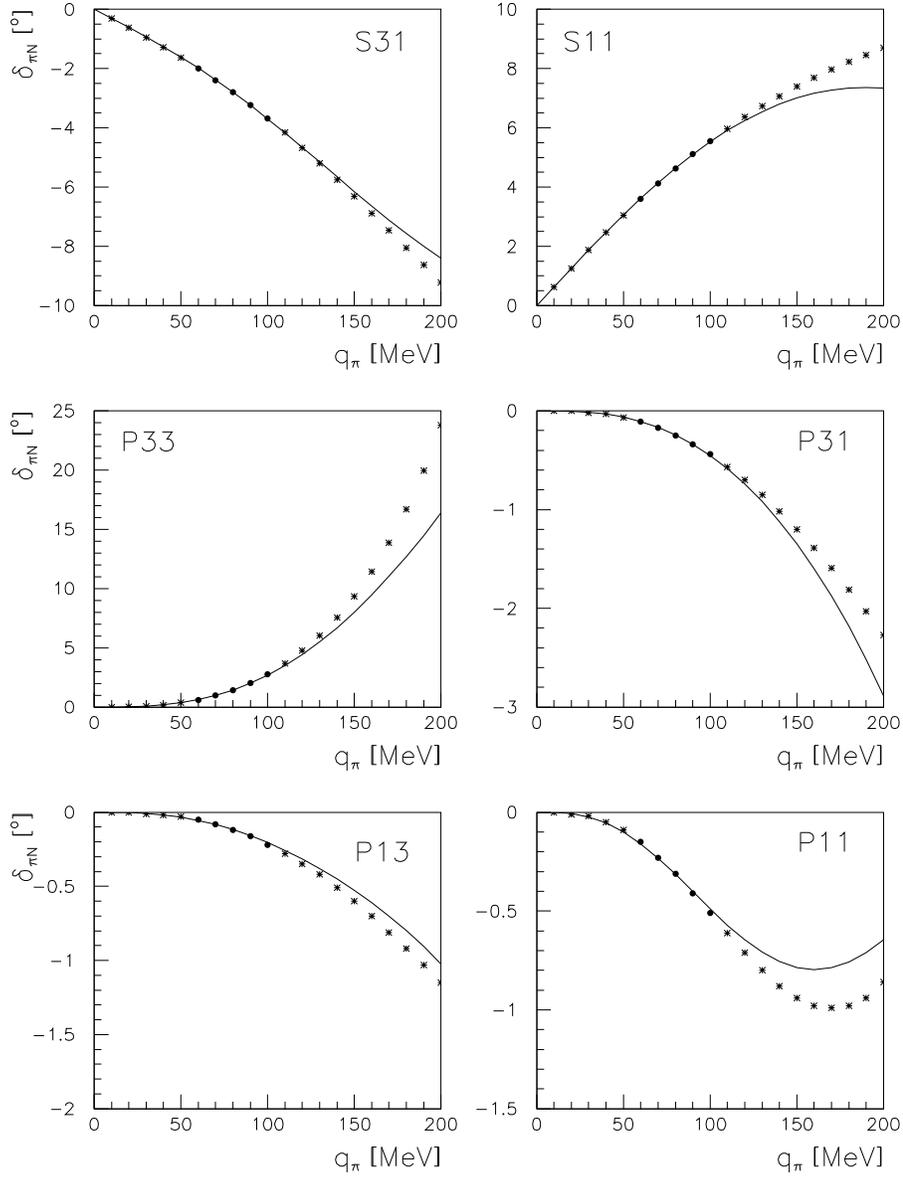}
}
\vskip 0.5cm
\caption{
Fits and predictions for the SP98 phase shifts as a
function of the pion laboratory momentum $q_\pi$. Fitted in
each partial wave are the data between 60 and 100~MeV (filled circles). For higher
and lower energies, the phases are predicted as shown by the solid lines.
}

\end{figure}

\vskip 1cm

\begin{figure}[bht]
\centerline{
\epsfysize=4in
\epsffile{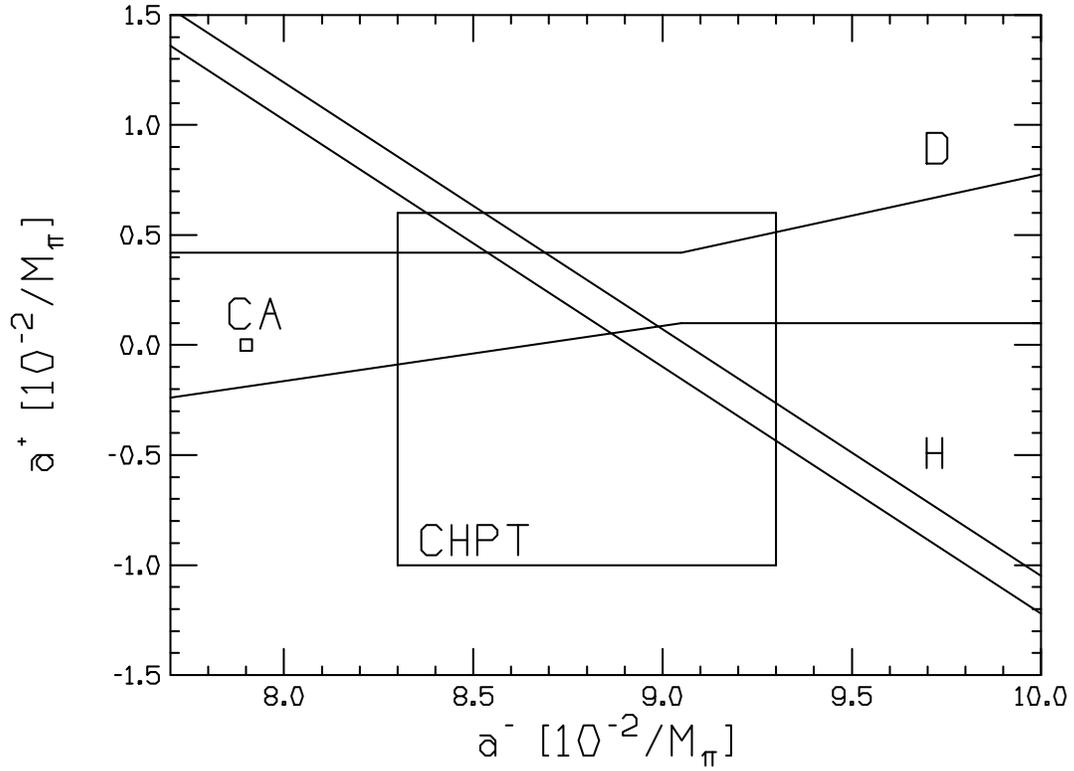}
}
\vskip 0.5cm
\caption{S--wave scattering lengths. Shown are the results obtained
from the level shift in pionic hydrogen (H--band) and pionic deuterium 
(D--band) compared to the
current algebra (CA)  and order $q^3$ CHPT prediction.
\label{fig:apm}}

\end{figure}

\vskip 1cm

\begin{figure}[bht]
\centerline{
\epsfysize=7in
\epsffile{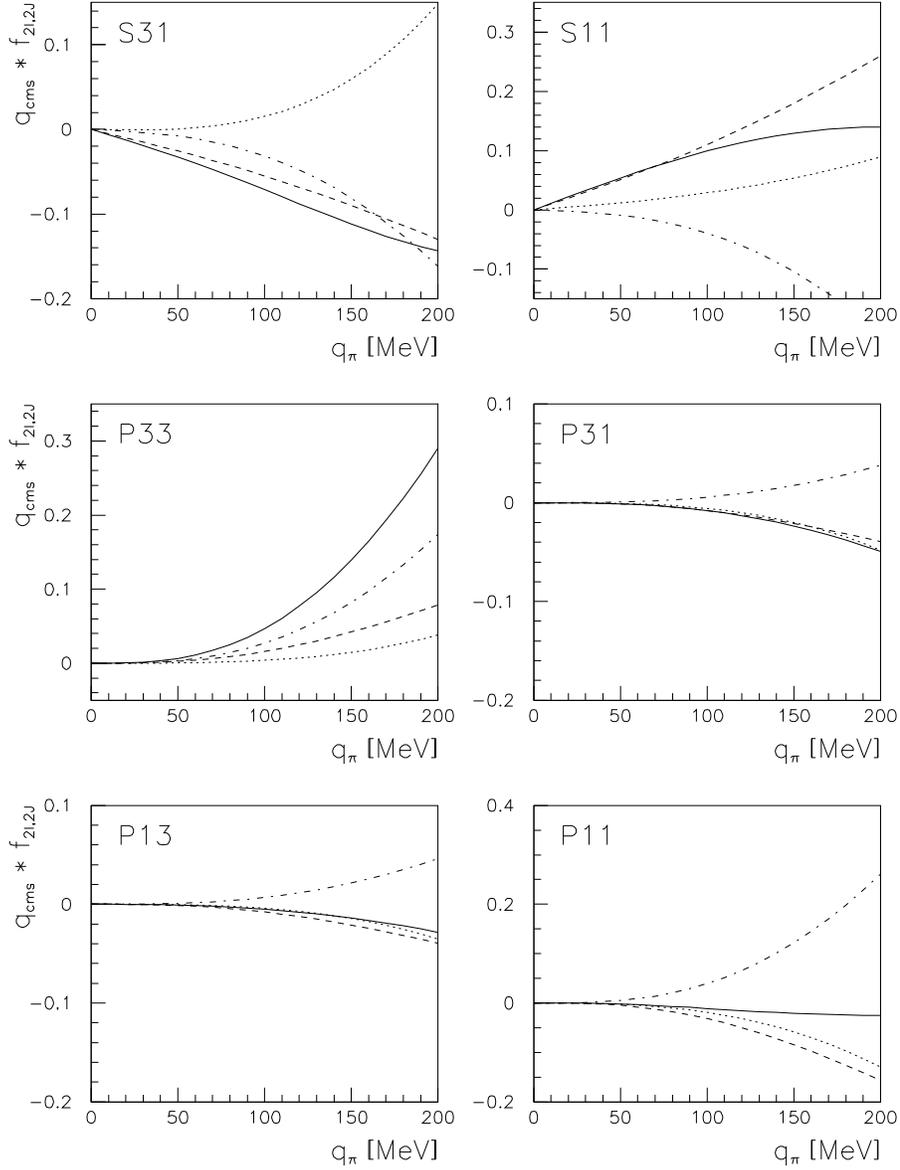}
}
\vskip 0.5cm
\caption{
Convergence properties as exemplified by the KA85 partial wave amplitudes.
Shown are the real part of the amplitudes multiplied by the modulus of the
pion cm momentum. The dashed, dashed--dotted and dotted lines give the first,
second and third order, respectively. The  sum is depicted by the solid lines.}

\end{figure}

\end{document}